\documentclass[onecolumn, twoside]{IEEEtran}

\usepackage[ruled]{algorithm2e}
\renewcommand{\algorithmcfname}{ALGORITHM}
\SetAlFnt{\small}
\SetAlCapFnt{\small}
\SetAlCapNameFnt{\small}
\SetAlCapHSkip{0pt}
\IncMargin{-\parindent}
\usepackage{graphicx}
\usepackage{algorithmic}
\usepackage{subfig}
\usepackage{multirow}
\usepackage{array}
\usepackage{amsmath} 
\usepackage{amssymb}
\usepackage[maxfloats=30]{morefloats}
\usepackage{amsthm}
\theoremstyle{definition}
\newtheorem{definition}{Definition}

\begin{document}
		\title{\huge Stable Desynchronization for Wireless Sensor Networks: \\(I) Concepts and Algorithms}

	\author{
		\IEEEauthorblockN{Supasate Choochaisri, Kittipat Apicharttrisorn, Chalermek Intanagonwiwat}\\
		\IEEEauthorblockA{Chulalongkorn University, Bangkok, Thailand\\Email:\{supasate.c, kittipat.api, intanago\}@gmail.com}
	}

	\date{}
	\maketitle

\begin{abstract}
Desynchronization is one of the primitive services for complex networks because it arranges nodes to take turns accessing a shared resource. TDMA is a practical application of desynchronization because it allows node to share a common medium. In this paper, we propose a novel desynchronization algorithm using \emph{artificial force field} called Multi-hop Desynchronization With an ARtificial Force field or M-DWARF and use it to solve TDMA problems in wireless sensor networks (WSNs). M-DWARF solves hidden terminal problems in multi-hop networks by using \emph{relative time relaying} and improves resource utilization by employing \emph{force absorption}. M-DWARF is suitable for use in WSNs because of its low complexity and autonomous operations. We implement M-DWARF using TinyOS \cite{tinyos} and experiment it on both TelosB motes \cite{telosb} and TOSSIM \cite{Levis:2003:TAS:958491.958506}. From our investigation, M-DWARF is the only desynchronization algorithm that achieves fast convergence with high stability, while maintaining channel utilization fairness. Moreover, we are the first to provide a stability analysis using dynamical systems and to prove that M-DWARF is stable at equilibrium.  (This paper is the first part of the series Stable Desynchronization for Wireless Sensor Networks - (I) Concepts and Algorithms (II) Performance Evaluation (III) Stability Analysis)

\end{abstract}

\section{Introduction}
Wireless sensor networks (WSNs) have attracted attention from wireless network research communities for more than a decade \cite{Akyildiz:2002:WSN:Survey}. Their abilities to collect environmental data and, to effectively and wirelessly send those data back to central, processing nodes, have been identified by research work such as \cite{senvm}. Recently, with the emerging Internet of Things (IoT) \cite{stankovic_iot}, researchers foresee its potentials to bring a new generation of the Internet where things are connected to engender unprecedented intelligence that facilitates people's daily lives, enterprise's tasks, and city-wide missions. WSNs are an integral part of the IoT because they are the sources of sensed data to be processed and analyzed by the computing clouds. Therefore, WSNs in the IoT are expected to handle operations with more density, diversity, and tighter integration.

WSNs often demand that sensed data be incorporated with timestamps so that the systems can fuse, distinguish and sequence the data consistently. Therefore, several protocols proposing to synchronize a global time in WSNs include \cite{Maroti:2004:FTS:1031495.1031501,5211944,Apicharttrisorn:2010:EGT:1900724.1901046}. However, some systems do not require a global notion of time; they demand sensor nodes to sample data at the same time to take sequential snapshots of an environment. Such systems include \cite{Werner-Allen:2005:FSN:1098918.1098934}, which is the first protocol to achieve this task of ``synchronicity''.  Desynchronization is an \emph{inverse} of synchronicity because it requires nodes \emph{not} to work at same time; hence, desynchronization can provide nodes with collision-free and even equitable access to a shared resource. A concrete example is a system using a Time Division Multiple Access or TDMA protocol, in which nodes utilize a shared medium at different time slots to avoid collision. In addition, desynchronization can schedule duty cycles of sensor nodes; in other words, nodes covering the same sensing area take turns waking up to sense the environment while others are scheduled to sleep to save the limited energy. Other potential applications of desynchronization include techniques to increase a sampling rate in multiple analog-to-digital converters, to schedule resources in multi-core processors, and to control traffic at intersections \cite{4274893}.

In this paper, we propose a stable desynchronization algorithm for multi-hop wireless sensor networks called Multi-hop Desynchronization With an ARtificial Force field or M-DWARF. We use TDMA to validate our algorithm and evaluate its performance. M-DWARF uses basic concepts of artificial force fields in DWARF \cite{Choochaisri:2012:DAF:2185376.2185378}. However, to support multi-hop networks and to avoid their hidden terminal problems, M-DWARF adds two mechanisms called \textit{Relative Time Relaying} and \textit{Force Absorption}. With these features added, M-DWARF is able to desynchronize multi-hop networks without collision from hidden terminals while maintaining maximum channel utilization. We evaluate M-DWARF's functionality on TelosB motes \cite{telosb} and its performance on TOSSIM \cite{Levis:2003:TAS:958491.958506}, a TinyOS simulator. We compare M-DWARF with Extended Desynchronization for Multi-hop Networks or EXTENDED-DESYNC (also referred as EXT-DESYNC) \cite{MK09DESYNC} and Lightweight coloring and desynchronization for networks or LIGHTWEIGHT \cite{5062165} on several topologies of multi-hop wireless networks. According to the simulation results, M-DWARF is the only desynchronization that has all the three properties - fast convergence, high stability, and maintained fairness. Moreover, in \cite{Choochaisri:2012:DAF:2185376.2185378}, we prove that desynchronization using artificial force fields is a convex function; that is, it converges to a global minimum without local ones. In addition, our stability analysis (in a supplementary document) proves that once DWARF or M-DWARF systems reach a state of desynchrony, they become stable. In other words, DWARF or M-DWARF provides a static equilibrium of desynchronized force fields at steady states. Once nodes in the systems deviate from a state of desynchrony, they will attempt to return back to the balance of the forces immediately. Our stability analysis not only suggests the stability of our desynchronization our algorithms but also proves that the systems will eventually converge to an equilibrium.

In the next section, we briefly explain how our proposed desynchronization, M-DWARF, works and describe its contributions. In Section \ref{sec:related_work}, we survey related literature of desynchronization in temporal and spatial domains, as well as TDMA scheduling algorithms. Section \ref{sec:desync_algo} recalls the basic concepts of DWARF and then explains the M-DWARF algorithm in detail. Finally, Appendix \ref{sec:psuedocode_m_dwarf} shows the psuedocodes of M-DWARF.

\section{Contributions}
\label{sec:contribution}
In order to understand the contributions of the M-DWARF algorithm, it is important to know basically how it works.
In addition to DWARF, M-DWARF has two mechanisms, which are named relative time relaying and force absorption, to support multi-hop topologies. The algorithmic steps of M-DWARF can be enumerated as follows.

\begin{enumerate}
\item Nodes, which are not initially desynchronized, set a timer to fire in $T$ time unit and listen to all one-hop neighbors.
\item Upon receiving a firing message from its one-hop neighbor, the node marks the current time to be the relative phase reference. Then, it reads relative phases of its two-hop neighbors which are included within the firing message. After that, it calculates their two-hop neighbors' phases by using the relative phase reference as the offset. The details are explained in Section \ref{sec:relative}.
\item When the timer, $T$, expires, a node broadcasts a firing message containing relative phases of its one-hop neighbors. Then, it calculates a new time phase to move on the phase circle, which is based on the summation of artificial forces from all phase neighbors within two hops where some forces are absorbed as explained in Section \ref{sec:absorption}. Then, the node sets a new timer according to the new calculated phase.
\item Similar to DWARF, a node adjusts its phase by using Eq. \ref{eq:newphase} with  $K$ calculated from Eq. \ref{eq:arbitrary_T}. 
\end{enumerate}  

All nodes in the artificial force field (in the period circle) iteratively run the same algorithm until the force is balanced.
The pseudo-code of this algorithm is shown in Appendix \ref{sec:psuedocode_m_dwarf}. The contributions of our proposed desynchronization algorithms are listed as follows.

\paragraph{Autonomous operations} M-DWARF works without any master or root nodes, so it does not need to elect or select such nodes and also does not have a single point of failure in this sense. Moreover, nodes do not need knowledge about network topology; they only have to know their one-hop and two-hop neighbors in order for the desynchronization to function correctly. In addition, M-DWARF is easy to be deployed because it works independently without any deployment setup. Finally, M-DWARF adapts itself very well to dynamics such as leaving or joining nodes.

\paragraph{Determinism} M-DWARF does not use any random operation, so its protocol behavior is deterministic and predictable.

\paragraph{Throughput} Thanks to its high stability, at a state of desynchrony, a node's time slot firmly stays at the same position in the next iteration, causing no or minimal interference with adjacent time slots. As a result, M-DWARF requires less \emph{guard time} between slots, allowing nodes to fully utilize the resource or medium because of its stable desynchronization. Without stability, the beginning and the end of the frames are likely to collide or interfere with the adjacent ones because time slots of the next iteration do not strictly stay at the same position. Moreover, M-DWARF requires low overhead. In each time period \textit{T}, each node only broadcasts a desynchronization message containing one-hop and two-hop neighbor information; in other words, it does not need any two-way or three-way handshakes like many other protocols.

\paragraph{Delay} A node that starts to join the desynchronization network needs to broadcast a desynchronization message to declare itself to the network and occupy its own time slot. Therefore, it has to wait one time period to send the first data frame. By determining the slot size and transmission bitrate, the node can predict how many iterations it needs to finish transmitting one packet. In consequence, all the nodes in the network route can share information regarding the end-to-end delay of a packet traversal. Therefore, upper and lower limits of such a delay can be determined.

\paragraph{Interoperability} Because M-DWARF does not assume any radio hardware specifications or MAC-layer techniques, it is highly interoperable with open standards. It assumes only that all nodes can transmit and receive data at the same spectrum. Although WSNs are a target platform of this paper, we argue that our desynchronization and TDMA algorithms can be applied to generic wireless multi-hop networks \cite{Sgora:2015:TDMA} or wireless mesh networks \cite{Vijayalayan:2013:Scheduling}.

\paragraph{Complexity} M-DWARF has low complexity for the following three reasons. First, it does not require time synchronization between nodes. Second, there are no handshakes; only one-way broadcasting is necessary. Third, its computational complexity depends only on the number of one-hop and two-hop neighbors, instead of the entire network size.

\paragraph{Energy Efficiency} Because of M-DWARF's high stability, data frames of adjacent time slots are less likely to collide or interfere with each other. Therefore, it has lower probability for packet retransmission, which is a factor for energy inefficiency in networks.

\paragraph{Channel Utilization Fairness} M-DWARF can provide equal access for all the neighbor nodes. The artificial forces are balanced between neighbor nodes, so all the nodes converge to occupy equal slot sizes.

\section{Related Work}
\label{sec:related_work}
Our related work can be mainly divided into three categories - 1) desynchronization on a temporal domain in wireless networks 2) desynchronization on a spatial domain in robotics 3) TDMA scheduling algorithms. Then, we summarize the properties of related work of the first category in Table \ref{tab:compared}.

\subsection{Desynchronization on a Temporal Domain in Wireless Networks}
\label{sec:timedesync}
To the best of our knowledge, Self-organizing desynchronization or DESYNC \cite{4379660}  is the first to introduce the desynchronization problem. In DESYNC, a node simply attempts to stay in the middle phase position between its previous and next phase neighbors. By repeating this simple algorithm, all nodes will eventually and evenly be spread out on a temporal ring. However, the error from one phase neighbor is also propagated to the other phase neighbors and is indefinitely circulated inside the network. As a consequence, DESYNC's error is quite high even after convergence. C. M. Lien et al. propose Anchored desynchronization or ANCHORED \cite{anchored} that uses the same method as DESYNC but requires one anchored node to fix the phase of its oscillator. However, because ANCHORED uses only the phase information of the phase neighbors, it shall suffer the similar desynchronization error as DESYNC. In contrast, our work relies on all received neighbors' information and is therefore more robust to the error from one phase neighbor. In  \cite{4663417}, the authors describe how DESYNC works on multi-hop networks and explain an extension for DESYNC by exchanging two-hop neighbor information. 

Inversed Mirollo-Strogatz or INVERSE-MS \cite{4274893}, designed to converge faster than DESYNC, is an inverse algorithm of the synchronicity work by \cite{MS1990}. 
At a steady state, INVERSE-MS maintains a dynamic equilibrium (\textit{i.e.}, nodes keep changing time positions while maintaining desynchronization). However, in INVERSE-MS, the time period is distorted whereas our algorithm does not distort the time period.
In Extended Desynchronization or EXT-DESYNC \cite{MK09DESYNC}, the authors propose a desynchronization algorithm that is similar to the extension proposed in \cite{4663417}. Each node sends its one-hop neighbors' relative time information to all of its one-hop neighbors.
Then, the one-hop neighbors relay such information to two-hop neighbors so that each node knows two-hop relative time information.
Consequently, each node can presume that there are two-hop neighbors appearing on the time circle.
Therefore, each node uses time information of both one-hop and two-hop neighbors to desynchronize with the same algorithm as in DESYNC. One mechanism in our multi-hop algorithm proposed in this paper is partly based on this notion.

M-DESYNC \cite{5062256} is a localized multi-hop desynchronization algorithm that works on a granularity of time slots. This protocol uses a graph coloring model for solving desynchronization. It starts by estimating the required number of time slots with a two-hop maximum degree or the maximum number of colors. This estimation allows nodes in M-DESYNC to immediately choose each predefined slot or color and helps M-DESYNC converge very fast. However, M-DESYNC requires that all nodes have a global notion of time in order to share the common perception of time slots. Furthermore, M-DESYNC claims that it works only on acyclic networks. On the contrary, our algorithm does not require a global notion of time and can work on both acyclic and cyclic networks.

A. Motskin et al. \cite{5062165} propose a simple, lightweight desynchronization algorithm, namely LIGHTWEIGHT, that is also based on a graph coloring model. Unlike M-DESYNC, the algorithm works on general graph networks and does not need the global time. To ensure that the selected time slot does not overlap with others', a node needs to listen to the shared medium for a full time period before claiming the slot. The listening mechanism can only avoid collision with one-hop neighbors but cannot avoid collision with two-hop neighbors (\textit{i.e.}, the hidden terminal problem). On the contrary, our algorithm works well on multi-hop networks; each node can effectively avoid collision with two-hop neighbors.
Furthermore, without a common notion of time, the starting time of each slot is quite random; as a result, several time gaps are too small to be used as time slots. This external fragmentation problem poorly reduces resource utilization of the system. Finally, to converge faster, their algorithm overestimates the number of required time slots. Hence, several large time gaps are also left unused and the resource utilization is undoubtedly and significantly lowered. In our work, nodes gradually adapt their phases to be separated from each other as far as possible. Therefore, the external fragmentation problem is reduced and the resource utilization is maximized. 

T. Pongpakdi et al. propose Orthodontics-inspired Desynchronization or DESYNC-ORT \cite{desyncort}. In their work, they use information from all one-hop neighbors and attempt to find nodes that are already in corrected time positions and tie them up together. This method is similar to the Orthodontics method that attempts to tie teeth which are already in corrected positions together.
The desynchronization errors of DESYNC-ORT are lower than those of DESYNC.
However, to calculate the correct positions, each node is required to know the total number of nodes in the system in advance. Additionally, the algorithm does not propose to solve the problem in multi-hop networks because nodes in two-hop neighbors cannot be tied together with one-hop neighbors. In contrast, our algorithm does not require nodes to have knowledge about the total number of nodes in advance but gradually and automatically adapts itself based on the current number of neighbors. Finally, our algorithm works on multi-hop networks. 

Vehicular-network Desynchronization, abbreviated as V-DESYNC \cite{v-desync}, is proposed to desynchronize nodes in vehicular ad-hoc networks. Their work has a different objective; that is, the algorithm does not focus on fairness (\textit{i.e.}, nodes are not necessary to be equitably separated) because vehicular networks are highly dynamic. In our work, we focus on wireless sensor networks with static sensor nodes and we attempt to render resource utilization fairness among sensor nodes.

Table \ref{tab:compared} summarizes the features of works in this category. Note that the overhead of the proposed algorithm depends on whether the algorithm works on the single-hop or multi-hop mode.

\begin{table*}[t]
	\centering
	\renewcommand{\arraystretch}{1.2}
\caption{Comparison of Desynchronization Protocols}
	{
		\begin{tabular}{|m{2cm}|m{1.5cm}|m{1.5cm}|m{1.2cm}|m{1.2cm}|m{1.5cm}|m{1.2cm}|m{1.5cm}|m{1cm}|}
			\hline 
			\multirow{2}{*}{Algorithms} & \multicolumn{8}{c|}{Properties}\\
			\cline{2-9} 
			& Period & Time sync & Fair Space & Multi- hop & Conver- gence & Error & Scalable & Over- head\\
			\hline 
			\hline 
			DESYNC & Fixed & No & Yes & No & Moderate & High & Poor & Zero\\
			\hline 
			ANCHORED & Fixed & No & Yes & No & Moderate & High & Poor & Zero\\
			\hline 
			INVERSE-MS & Distorted & No & Yes & No & Fast & Low & Good & Zero\\
			\hline 
			EXT-DESYNC & Fixed & No & Yes & Yes & Moderate & High & Poor & High\\
			\hline 
			M-DESYNC & Fixed & Required & No & Yes & Fast & High & Good & Low\\
			\hline 
			LIGHT- WEIGHT & Fixed & No & No & Yes & Fast & High & Good & Zero\\
			\hline 
			DESYNC-ORT & Fixed & No & Yes & No & Moderate & Low & Good & Zero\\
			\hline 
			V-DESYNC & Fixed & No & No & No & No & High & Good & Zero\\
			\hline 
			DWARF & Fixed & No & Yes & No & Moderate & Low & Good & Very Low\\
			\hline 
			M-DWARF (Proposed) & Fixed & No & Yes & Yes & Moderate & Low & Good & Low\\
			\hline 
		\end{tabular}
	}
	
	\label{tab:compared}
\end{table*}

\subsection{Desynchronization on a Spatial Domain in Robotics}
\label{sec:spacedesync}

\begin{figure}
	\centering
	\subfloat[Robotic Close Ring]{
		\label{fig:subfig:robotic_closed_ring}
		\includegraphics[width=1.2in]{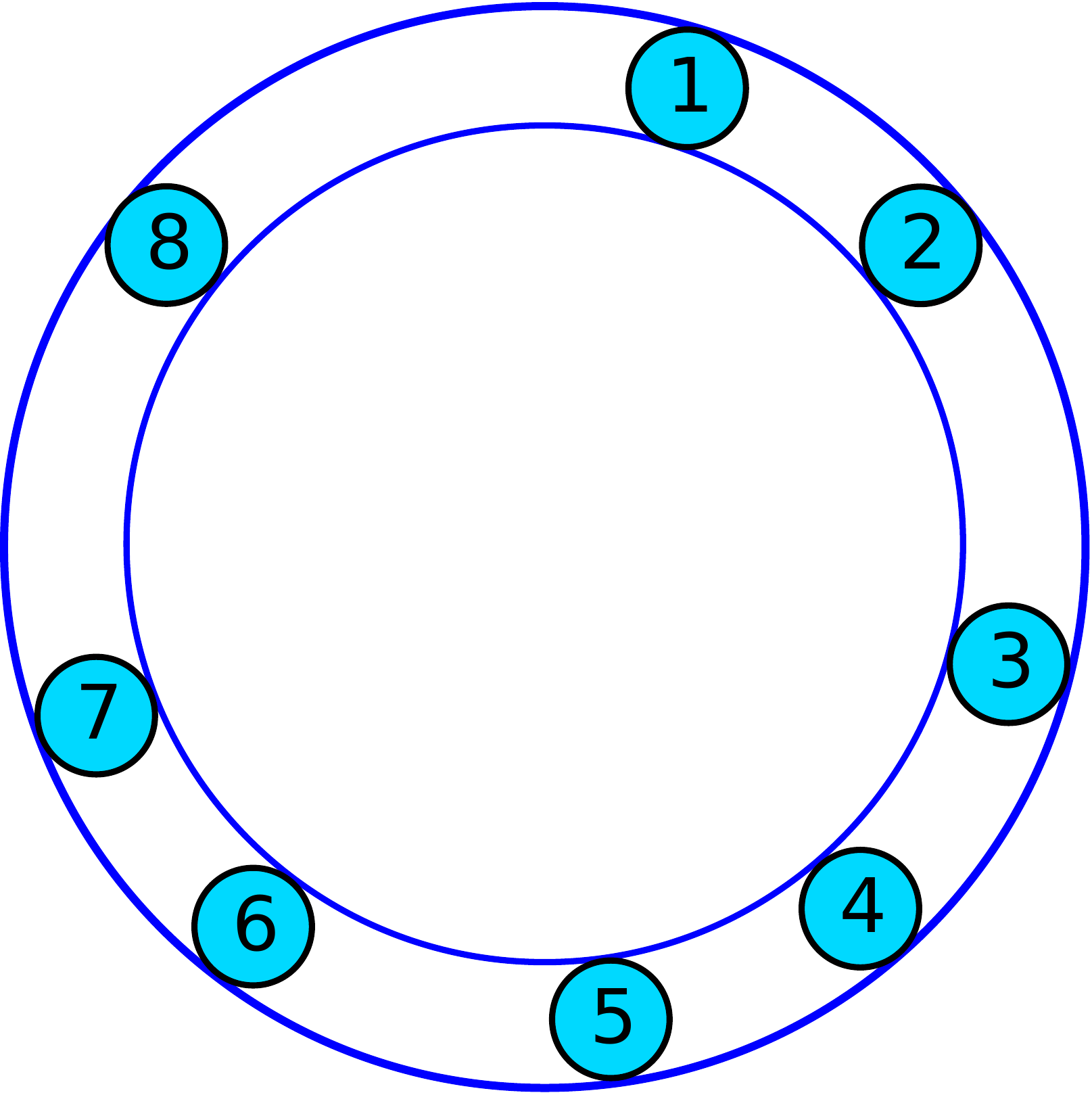}
	}
	\hspace{1.5cm}
	\subfloat[Robotic Perfect Close Ring]{
		\label{fig:subfig:robotic_closed_ring_perfect}
		\includegraphics[width=1.2in]{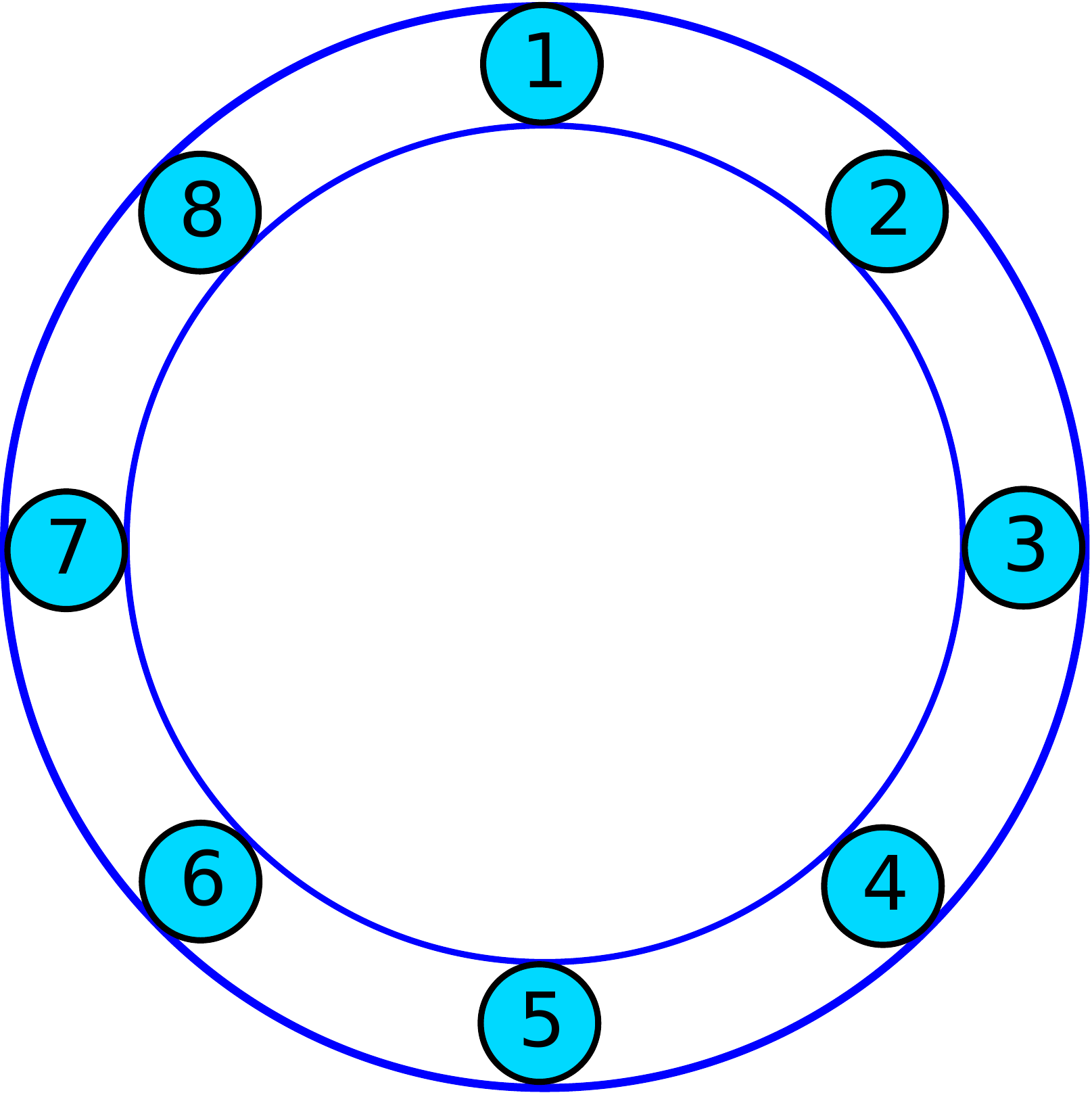}
	}
	\caption{Robotic pattern formation on a closed ring. (\ref{fig:subfig:robotic_closed_ring}) Robots are randomly placed on a closed ring. (\ref{fig:subfig:robotic_closed_ring_perfect}) In the perfect configuration, robots are equitably separated from each other.}
	\label{fig:robotic_ring}
\end{figure}

In robotic pattern formation, multiple robots distributedly group and align themselves in geometric patterns such as circles, rectangles, and triangles. Without an explicit argument, robotic pattern formation can be abstracted as desynchronization on a spatial domain. Robots attempt to separate away from each other as far as possible to form such patterns; in other words, robots desynchronize themselves spatially to avoid the collision with each other in the spatial domain.

These two pieces of work  \cite{suzuki-96,suzuki-99} investigate several robotic pattern formations. However, the pattern formation that is similar to desynchronization on the temporal domain is the formation on a closed ring. 
Figure \ref{fig:robotic_ring} illustrates a robotic formation on a closed ring. In Figure \ref{fig:subfig:robotic_closed_ring}, initially, robots are randomly placed on any position on the closed ring. The perfect configuration of the formation is illustrated in Figure \ref{fig:subfig:robotic_closed_ring_perfect}; that is, robots are equivalently separated away on the ring. 

\begin{figure}
	\centering
	\subfloat[A Robotic Move]{
		\label{fig:closedring-desync}
		\includegraphics[width=1.2in]{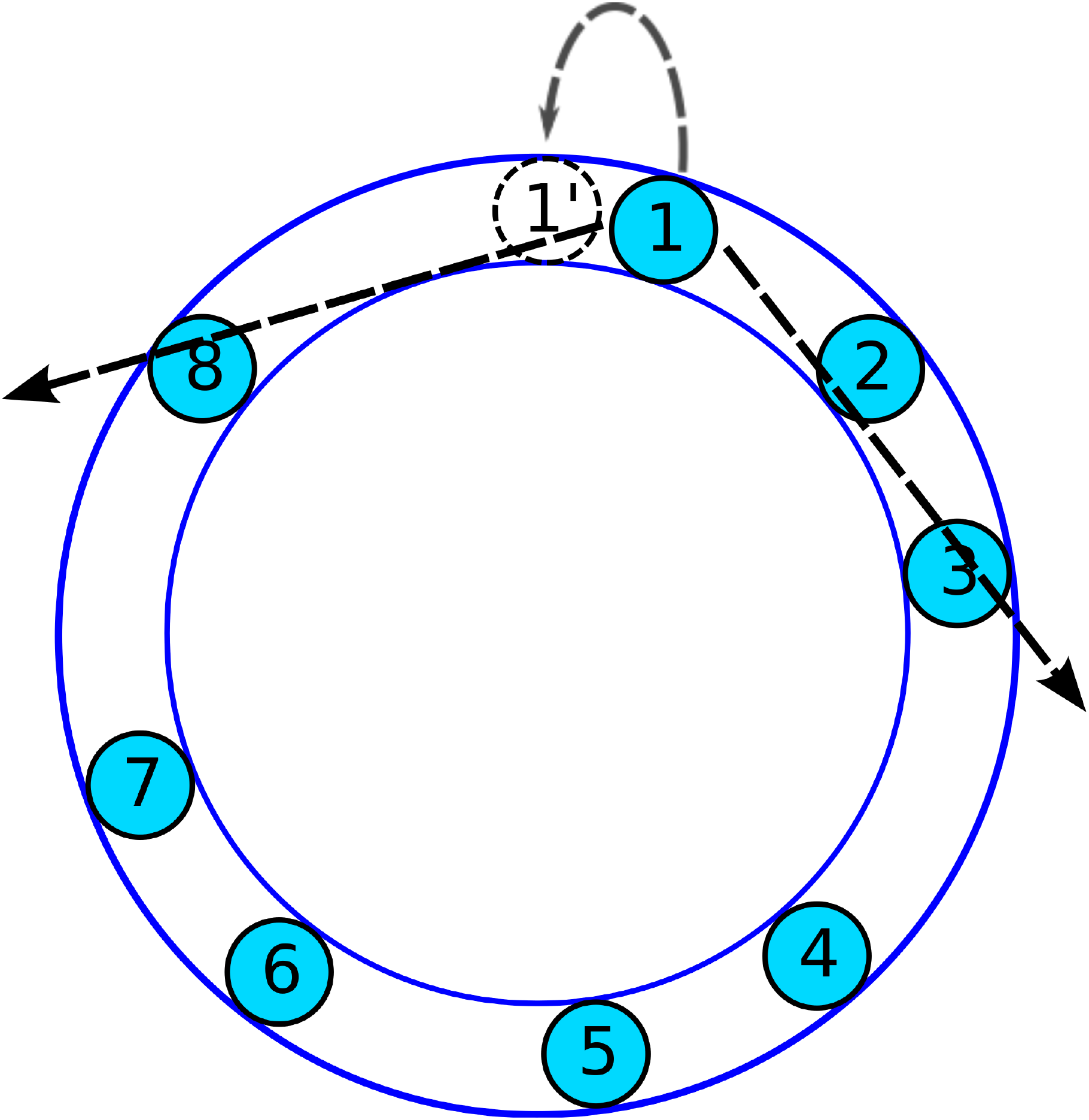}
	}
	\hspace{1.5cm}
	\subfloat[Convergence to Perfect Configuration]{
		\label{fig:closedring-desync-perfect}
		\includegraphics[width=1.2in]{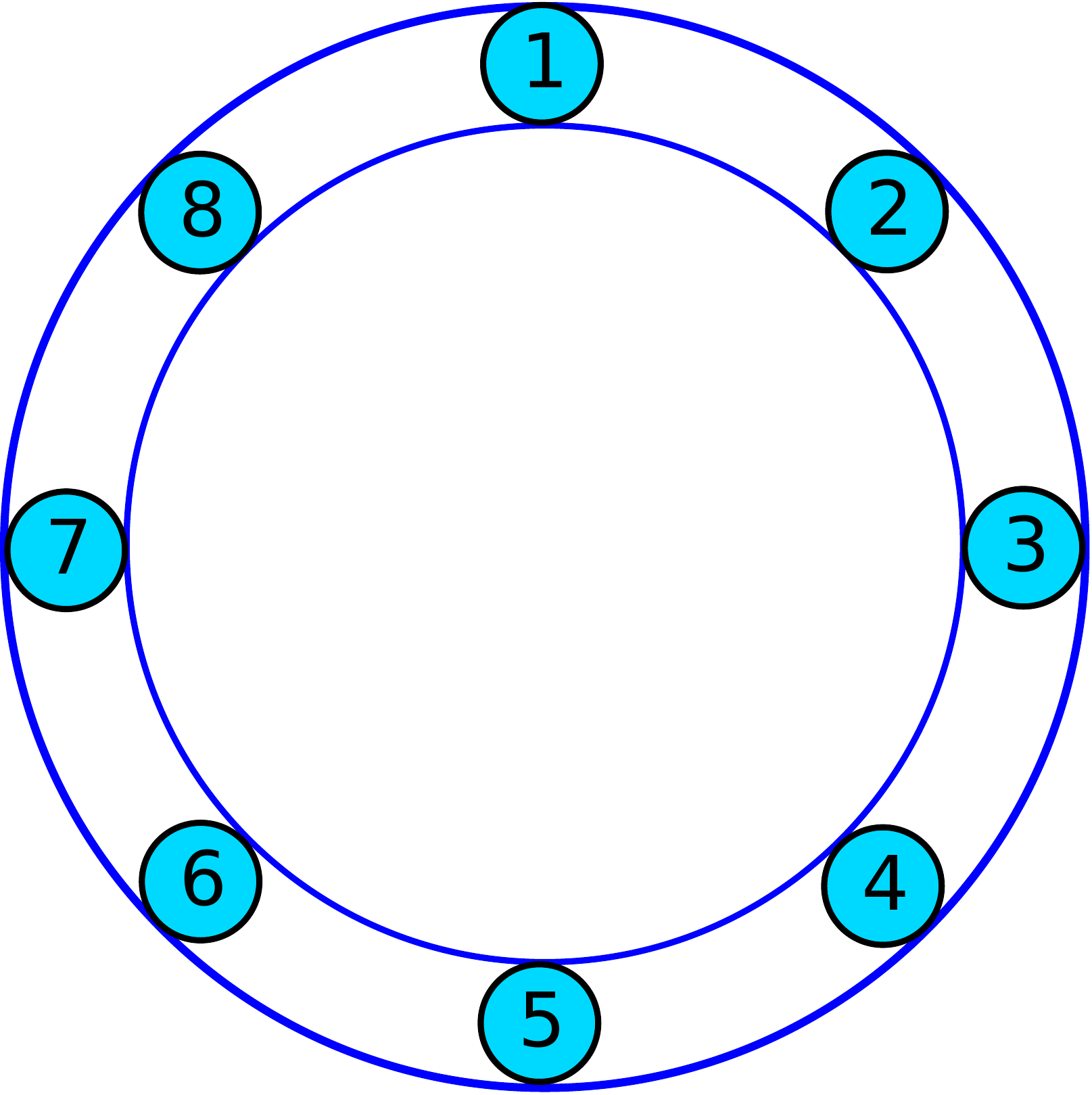}
	}
	\caption{Moving to the midpoint algorithm. (a) Each robot moves to the midpoint between two nearest visible neighbors. (b) The algorithm converges to the perfect configuration..}
	\label{fig:robotic-closed-ring-desync}
\end{figure}

Other papers such as \cite{defago04,cohen-08,flocchini-08} propose the algorithms that are similar to each other for robotic formations on a closed ring, assuming that robots have limited visibility range. Each robot attempts to adjust its position to the middle between two nearest robots on its left side and right side (Figure \ref{fig:closedring-desync}). In their papers, they prove that this simple algorithm eventually converges the robotic formation to the perfect configuration (Figure \ref{fig:closedring-desync-perfect}).

\begin{figure}
	\centering
	\includegraphics[width=3.0in]{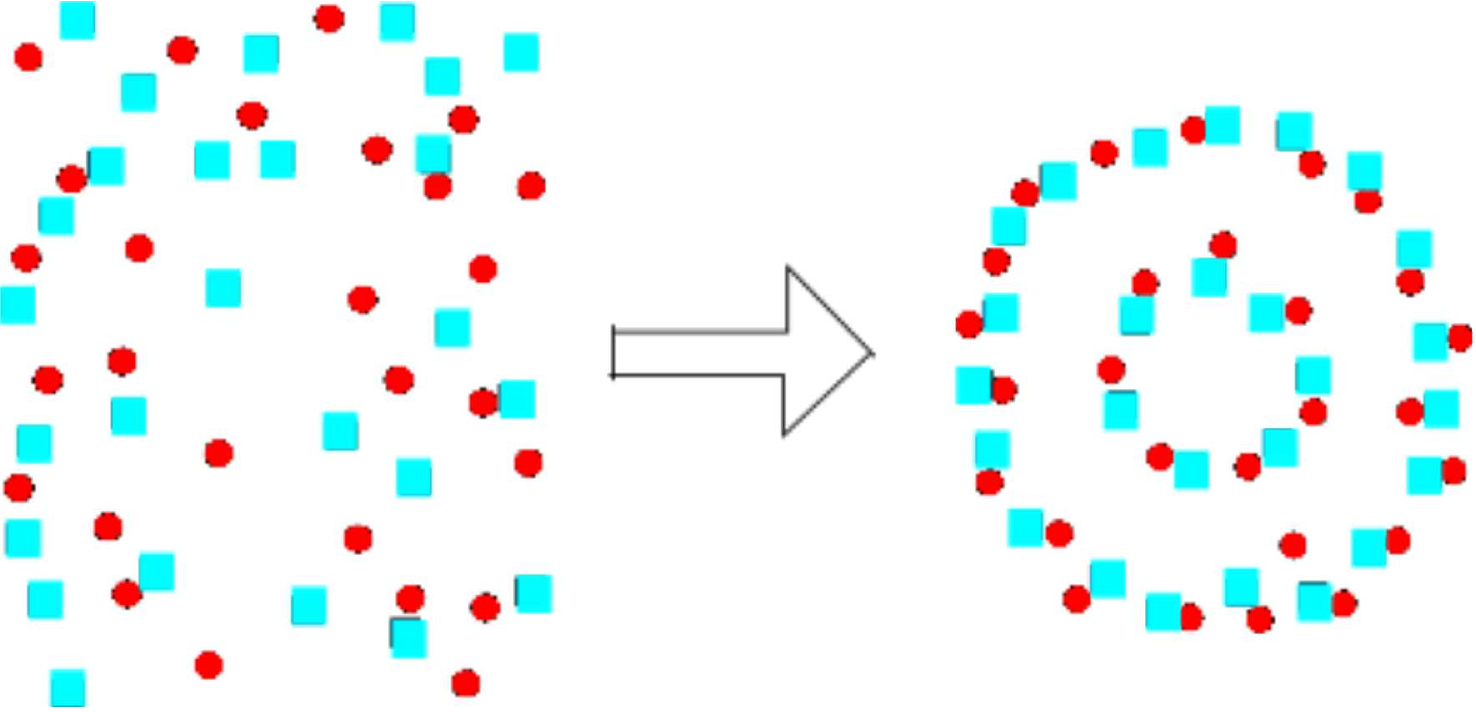}
	\caption{Results of Robotic Circular Formation. Robots with two different types form the circle.}
	\label{fig:circular_formation}
\end{figure}

In \cite{4141997}, heterogeneous robots are grouped in a distributed manner into teams that are equally spread out to cover the monitored area. Each robot has no global knowledge of others' absolute positions but can detect relative positions of the others with respect to itself as well as the type of the others. 
To form a circle, an artificial force is used as an abstraction for velocity adaptation of a robot. 
Robots of different types have attractive forces to each other, while robots of the same type have repulsive forces. As a result, a circle of heterogeneous robots is formed and robots are spaced on the circle  (see Figure \ref{fig:circular_formation}). This work inspires our desynchronization algorithm.

\subsection{TDMA Scheduling Algorithms}
\label{sec:tdma}

Other works that are related to desynchronization protocols are distributed Time Division Multiple Access (TDMA) protocols. Distributed TDMA protocols are similar to M-DESYNC \cite{5062256}; their protocols work on a granularity of time slots. Similar to M-DESYNC, many of the distributed TDMA protocols such as TRAMA \cite{trama}, Parthasarathy \cite{Parthasarathy}, ED-TDMA \cite{edtdma}, and Herman \cite{herman} assume time is already slotted or all nodes are synchronized to achieve the same global clock. 
In our work, we do not require time synchronization and do not assume already slotted time. 

S. C. Ergen et al. \cite{Ergen:2010:TDMAAlgo} propose node-based and level-based TDMA scheduling algorithms for WSNs. Their techniques mainly derive from graph coloring algorithms in which the \textit{colors} must have been predefined. In contrast, our desynchronization algorithms never predefine slots but rather allow nodes to adjust their slots with those in the neighborhood on-the-fly.
K. S. Vijayalayan et al. \cite{Vijayalayan:2013:Scheduling} survey distributed scheduling techniques for wireless mesh networks and A. Sgora et al. \cite{Sgora:2015:TDMA} provide an extensive survey of recent advances in TDMA algorithms in wireless multi-hop networks.
 %Distributed Scheduling Schemes for Wireless Mesh Networks: A Survey
Both the survey papers interestingly give a comprehensive overview of the TDMA scheduling algorithms and techniques that are still being investigated and further developed by wireless network researchers worldwide.

\section{DWARF and M-DWARF Desynchronization Algorithms}
\label{sec:desync_algo}
In this section, we briefly introduce the concept of \textit{artificial force field} and our previous work of desynchronization for single-hop networks \cite{Choochaisri:2012:DAF:2185376.2185378} because it is necessary to understand the basic concepts before the proposed algorithm for multi-hop networks can be understood. 

\subsection{Desynchronization Framework, Artificial Force Field and DWARF Algorithms}
\label{AFF_DWARF}
\subsubsection{Desynchronization Framework}
\begin{figure}
	\centering
	\includegraphics[width=3.2in]{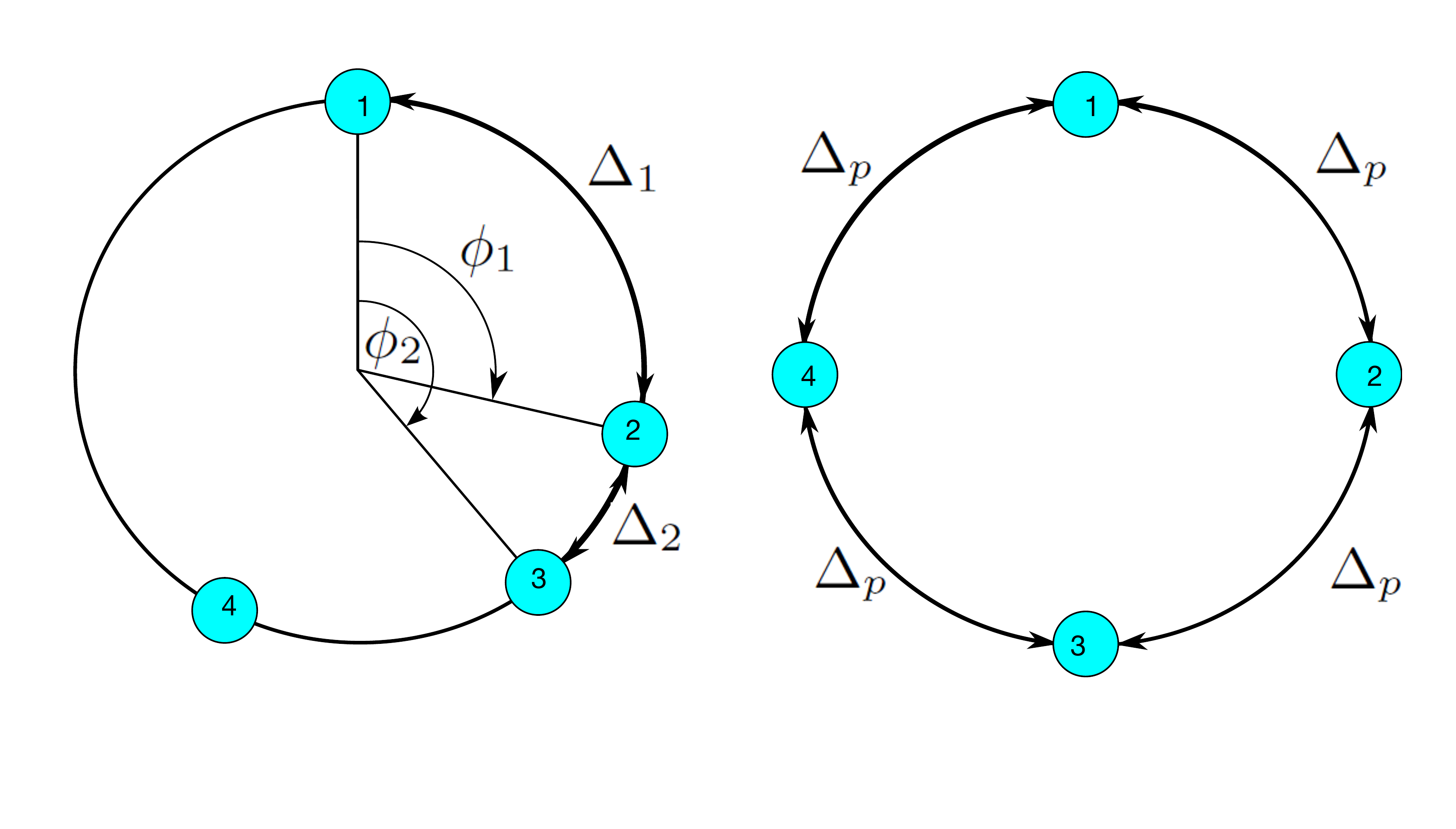}
	\caption{Desynchronization framework: $\phi_1$ and $\phi_2$ are phases of node 1 and 2 respectively. While all the four nodes are phase neighbors to each other, node 2 and 4 are the previous and next phase neighbor of node 1 respectively. The left figure shows a desynchrony state that will converge to the perfect desynchrony state as in the right figure.}
	\label{fig:time_circle}
\end{figure}

The desynchronization framework is depicted as a time circle in Figure \ref{fig:time_circle}.
The perimeter of a time circle represents a configurable time period $T$ of nodes' oscillators.
The time position or \textit{phase} of each node represents its turn to perform a task (\textit{e.g.}, accessing a shared resource, sampling data, or firing a message).
The system is desynchronized when all nodes are separated in the time circle. We define the terms used in the desynchronization context as follows.

\begin{definition}[Phase]
	A phase $\phi_i$ of node $i$ is the time position on the circle of a time period $T$, where $0 \leq \phi_i <  T$ and $T \in \mathbb{R}^+$.
\end{definition}

\begin{definition}[Phase Neighbor]
	Node $j$ is a phase neighbor of node $i$ if node $i$ perceives the existence of node $j$ through reception of $j$'s messages at the phase $\phi_i + \phi_{i,j}$, where $\phi_{i,j}$ is the phase difference between node $j$ and node $i$,
	\begin{equation}
	\phi_{i,j} = \left\{ 
	\begin{array}{l l}
	\phi_j - \phi_i & \quad \mbox{if $\phi_j \geq \phi_i$,}\\
	T - (\phi_i - \phi_j) & \quad \mbox{if $\phi_j < \phi_i$.}\\ \end{array} \right. 
	\end{equation}
\end{definition}

\begin{definition}[Next Phase Neighbor]
	Node $j$ is the next phase neighbor of node $i$ if $\phi_{i,j} = \underset{k \in S}{\min}\{\phi_{i,k}\}$, where $S$ is a set of node $i$'s neighbors. 
	%Node $j$ is the next phase neighbor of node $i$ if it has the minimum phase difference among node $i$'s phase neighbors . 
\end{definition}
\begin{definition}[Previous Phase Neighbor]
	Node $j$ is the previous phase neighbor of node $i$ if \\$\phi_{i,j} = \underset{k \in S}{\max}\{\phi_{i,k}\}$, where $S$ is a set of node $i$'s neighbors.
	%A node $j$ is the next phase neighbor of a node $i$ if it has the maximum phase difference among node $i$'s phase neighbors . 
\end{definition}
\begin{definition}[Desynchrony State]
	The system is in a desynchrony state if $\phi_i \neq \phi_j$ for all $i, j \in V$ and  $i \neq j$, where $V$ is a set of nodes in a network that cannot share the same phase.
\end{definition}
\begin{definition}[Perfect Desynchrony State]
	The system is in the perfect desynchrony state if it is in the desynchrony state and $\phi_{i,j} = T/N$ for all $i \in V$, $j$ is $i$'s previous phase neighbor, and $N$ is the number of nodes in a network that cannot share the same phase.
\end{definition}

We note that two nodes can share the same phase if they are not within the two-hop communication range of each other.

\subsubsection{Artificial Force Field}
\label{AFF}
An artificial force field is an analogy to the circle of a time period where nodes have repulsive forces to each other.
Nodes are in the same force field if they can communicate with each other or share the same medium. 
If node $i$ and node $j$ are on the same force field, they have repulsive forces to push one another away. 
A closer pair of nodes has a higher magnitude of force than a farther pair does.
The time interval between two nodes is derived from the phase difference between them.
If two nodes have a small phase difference, they have a high magnitude of force and vice versa.
In other words, a repulsive force is an inverse of a phase difference between two nodes:

\begin{equation}
f_{i,j} = - \frac{1}{\Delta \phi_{i,j} / T}, \quad \Delta \phi_{i,j} \in (-\frac{T}{2}, \frac{T}{2}),
\label{eq:force}
\end{equation}
where $f_{i,j}$ is the repulsive force from node $j$ to node $i$ on a time period $T$ and $\Delta \phi_{i,j}$ is the phase difference between node $i$ and $j$.
We note that $\Delta \phi_{i,j}$ is not equal to 0 because if two nodes fire at the same time, their firings collide and two nodes do not record other's firing. Additionally, at $T/2$ or $-T/2$, a node does not repel an opposite node because the forces are balanced.

\begin{figure}
	\centering
	\includegraphics[width=1.5in]{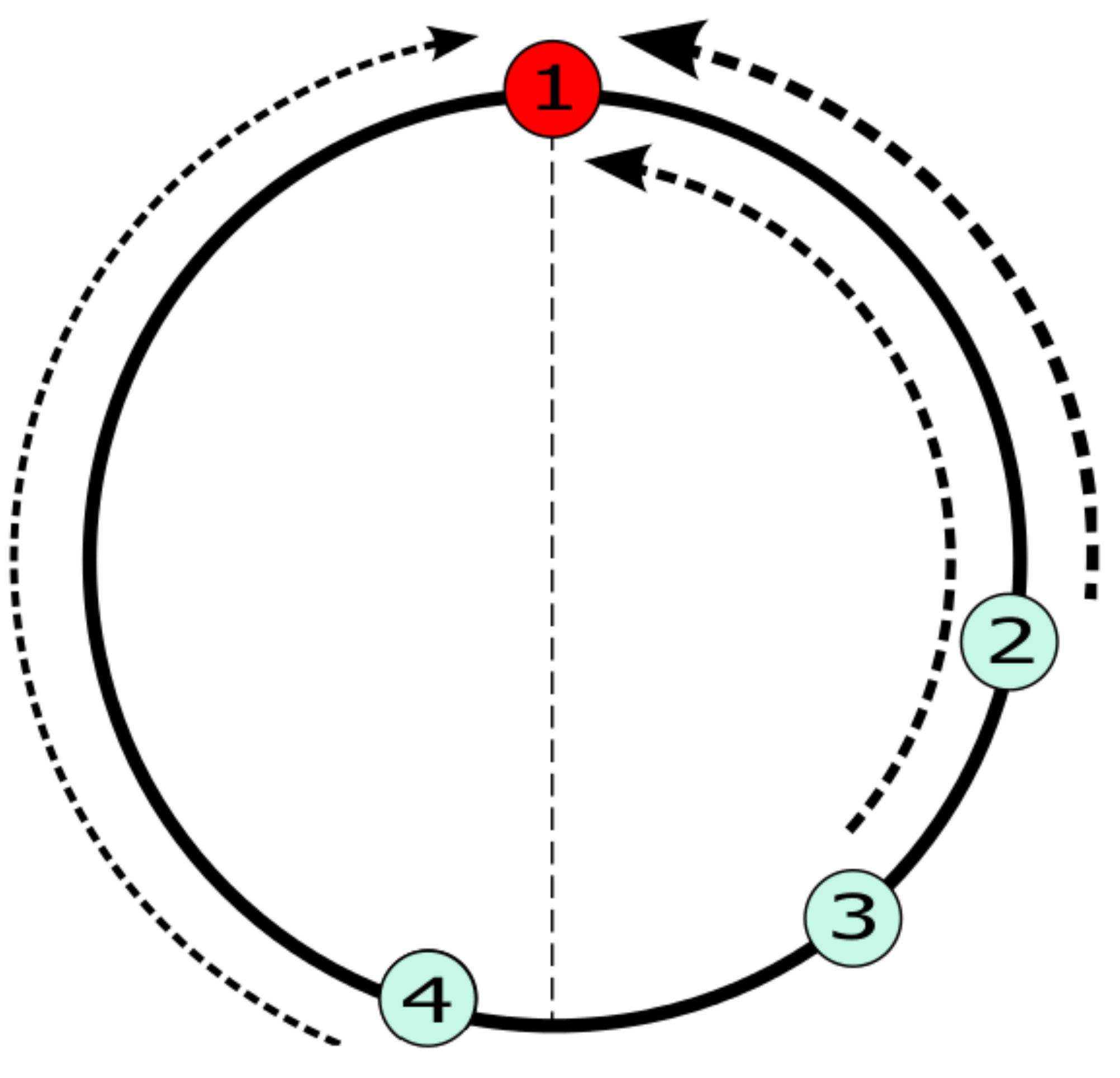}
	\caption{Artificial Force Field. Arrow lines represent repulsive forces from node 2, 3, and 4 to node 1. A shorter and thicker line is a stronger force. A force from node 4 is a positive force and two forces from node 2 and 3 are negative forces.}
	\label{fig:force_field}
\end{figure}

A repulsive force can be positive (clockwise repulsion) or negative (counterclockwise repulsion).
A positive force is created by a node on the left half of the circle relative to the node being considered whereas a negative force is created by a node on the right half.
Figure \ref{fig:force_field} represents a field of repulsive forces on node 1. 

Each node in the force field moves to a new time position or phase proportional to the total received force.
Given $n$ nodes in a force field, the total force on a node $i$ is the following:
\begin{equation}
\mathcal{F}_i = \sum_{\substack{j=1\\ j \neq i}}^{n}{f_{i,j}}.
\label{eq:fsum}
\end{equation}

Eventually, nodes reach an equilibrium where the total force of the system is close to zero and each pair of phase neighbor nodes has the same time interval.
This equilibrium state also indicates the perfect desynchrony state because all nodes are equally spaced on the time circle.
\subsubsection{DWARF, the Single-Hop Desynchronization Algorithm}
\label{DWARF}
We assume that, initially, nodes are not desynchronized and each node sets a timer to fire in $T$ time unit.
After setting the timer, each node listens to all its neighbors until the timer expires.
When receiving a firing message from its neighbor, the (positive or negative) repulsive force from that neighbor is calculated based on the phase difference.
When the timer expires, a node broadcasts a firing message to neighbors. 
Then, the node calculates a new time phase to move on the circle based on the summation of forces from all neighbors and sets a new timer according to the new time phase.

It is reasonable now to question how far a node should move or adjust its phase.
In our work, given the total received force $\mathcal{F}_i$, the node $i$ adjusts to a new time phase $\phi_i^{'}$,
\begin{equation}
\phi_i^{'} = (\phi_i + K\mathcal{F}_i) \mod T,
\label{eq:newphase}
\end{equation} 
where $\phi_i$ is the current phase of the node $i$.

Undoubtedly, the proper value of the coefficient $K$ leads to the proper new phase.
The value of $K$ is similar to a step size which is used in artificial intelligence techniques. 
Therefore, if the value of $K$ is too small, the system takes much time to converge. 
On the other hand, if the value of $K$ is too large, the system may overshoot the optimal value and fail to converge. 
We observe that, given the same time period, fewer nodes in the system result in bigger phase difference between two phase neighbors. To be desynchronized, nodes in sparse networks must make a larger adjustment to their time phases than nodes in dense networks.
Therefore, the same total received force should have a greater impact on a node in sparse networks than on a node in dense networks. 
To reflect this observation, the coefficient $K$ is inversely proportional to a power of the number of nodes $n$,
\begin{equation}
K = c_1 \times n^{-c_2}, \text{ where } c_1, c_2 \geq 0.
\end{equation}

Therefore, we have conducted an experiment to find the proper value of $c_1$ and $c_2$. 

\begin{figure}
	\centering
	\includegraphics[width=2.2in]{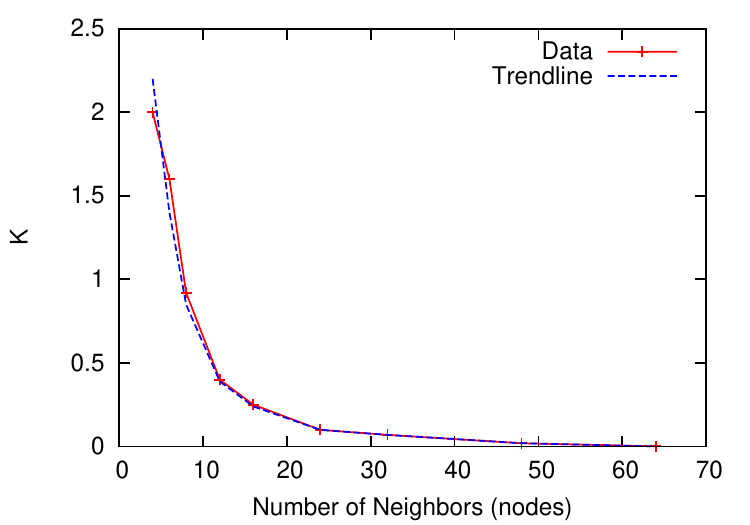}
	\caption{Relation of the coefficient $K$ with a number of nodes $n$}
	\label{fig:relation_k_n}
\end{figure}

We set a time period $T$ to 1000 and vary the number of nodes.
In the specific number of nodes, we first simulate to see the trend of the value $K$ that leads to small errors. 
Then, we select a range of good $K$ values and simulate 100 times to obtain the average desynchronization error for each $K$ value. 
In each simulation, we randomly set an initial phase of each node between 0 and $T$ (period value). 
Finally, we select the $K$ value that results in the lowest error. 
After getting the proper $K$ value for each number of nodes, we plot the relation between $K$ and the number of nodes (Figure \ref{fig:relation_k_n}) and use a mathematical tool to calculate the power regression. The obtained relation function between $K$ and $n$ (the trendline in Figure \ref{fig:relation_k_n}) consists of $c_1$ and $c_2$  values as follows:
\begin{equation}
K = 38.597 \times n^{-1.874}.  \nonumber
\end{equation}
However, this $K$ value is derived by setting $T$ equal to 1000. 
Therefore, for an arbitrary value of $T$, 
\begin{equation}
\label{eq:arbitrary_T}
K = 38.597 \times n^{-1.874} \times \frac{T}{1000}.
\end{equation}
The proof of Eq. \ref{eq:arbitrary_T} can be found in \cite{Choochaisri:2012:DAF:2185376.2185378}. Moreover, in \cite{Choochaisri:2012:DAF:2185376.2185378}, we also prove that the force function of DWARF has the convexity property; that is, it has one global minima and no local minima. Additionally, in this paper, we provide stability analysis of DWARF and M-DWARF in the Supplementary Material.

\subsection{M-DWARF, the Multi-hop Desynchronization Algorithm (Proposed)}
\label{sec:m_dwarf}
In this section, we extend the artificial force field concept to desynchronization in multi-hop networks. We begin with applying DWARF directly to a simple multi-hop topology and find out how the algorithm fails on such a topology in Section \ref{sec:hidden-terminal}. Then, we propose two simple yet effective resolutions, relative time relaying and force absorption, to extend DWARF for multi-hop networks in Section \ref{sec:relative} and \ref{sec:absorption} respectively. Additionally, we provide pseudo-code of M-DWARF in Appendix \ref{sec:psuedocode_m_dwarf}.

\subsubsection{The Hidden Terminal Problem}
\label{sec:hidden-terminal}
\begin{figure}
	\centering
	\includegraphics[width=2.5in]{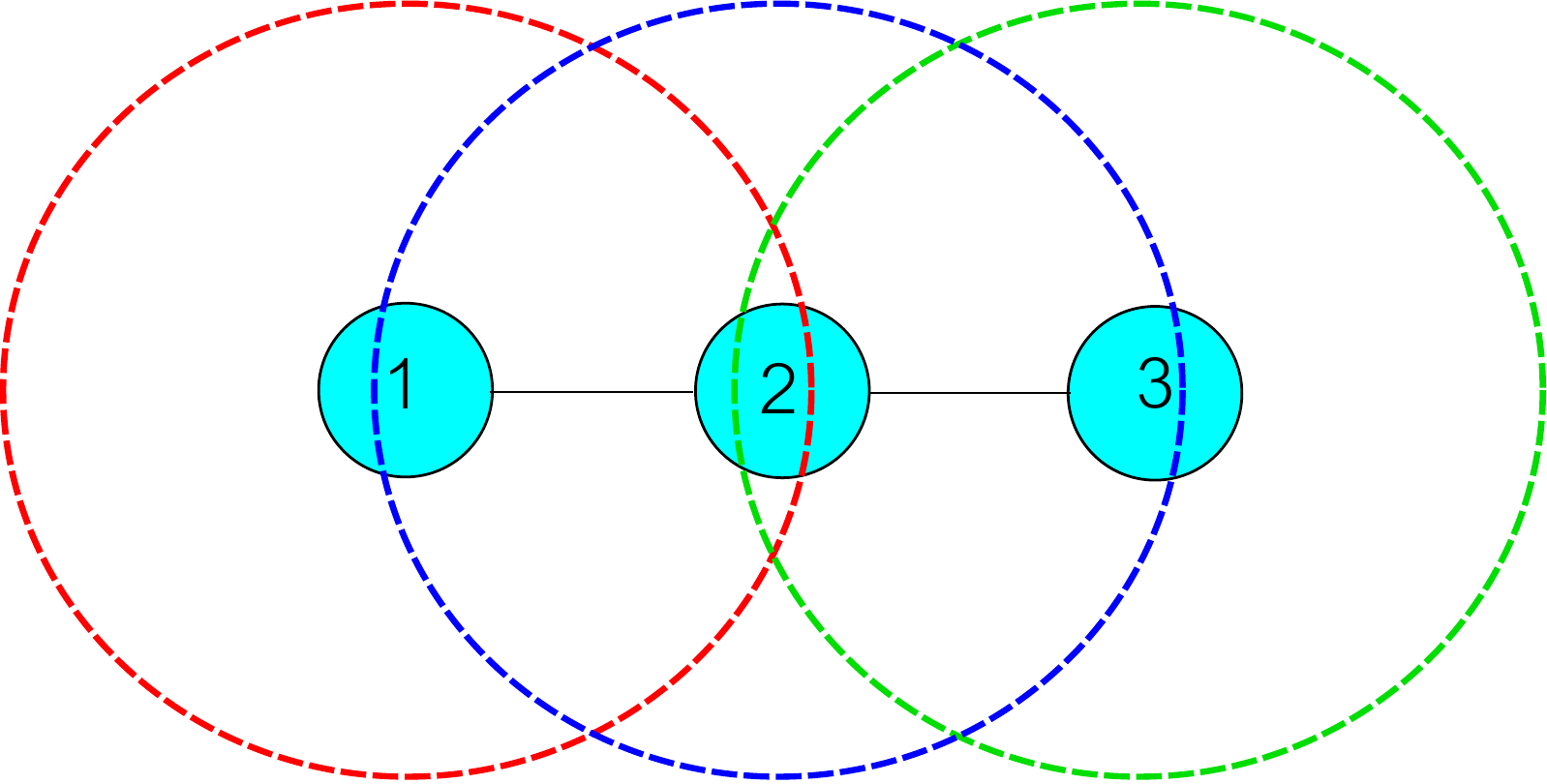}
	\caption{The hidden terminal problem.}
	\label{fig:3nodes-chain-hidden}
\end{figure}

To see how DWARF works on a multi-hop network, we set a simple 3-node chain topology as illustrated in Figure \ref{fig:3nodes-chain-hidden}. In this topology, node 1 can receive firing messages from node 2 but cannot receive from node 3; on the other hand, node 3 can receive firing messages from node 2 but cannot receive from node 1. However, node 2 can receive firing messages from both node 1 and 3 which are not aware of each other's transmission. This simple scenario causes messages to collide at node 2.

\begin{figure}
	\centering
	\subfloat[]{
		\includegraphics[width=2in]{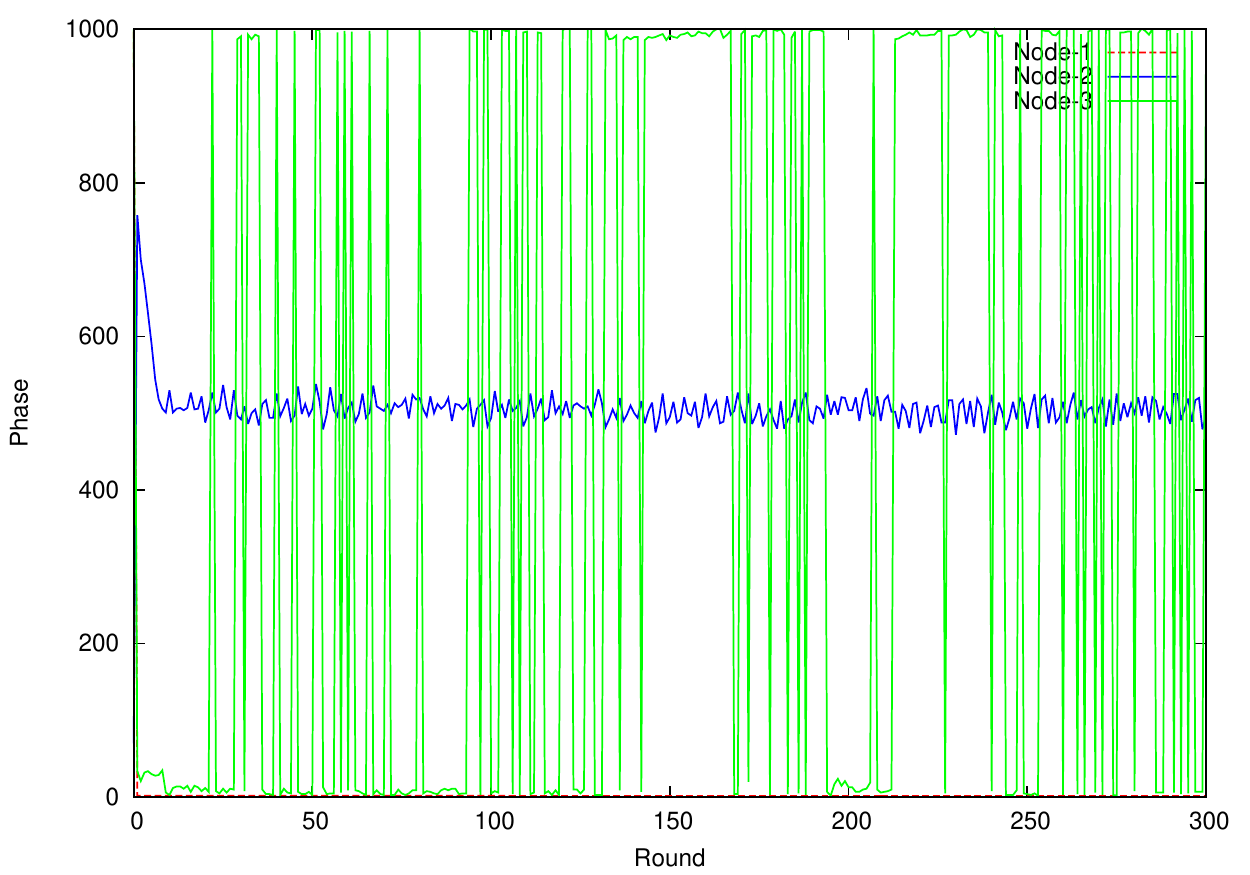}
		\label{fig:3nodes-chain-dwarf}
	}
	\hspace{1.0cm}
	\subfloat[]{
		\includegraphics[width=2in]{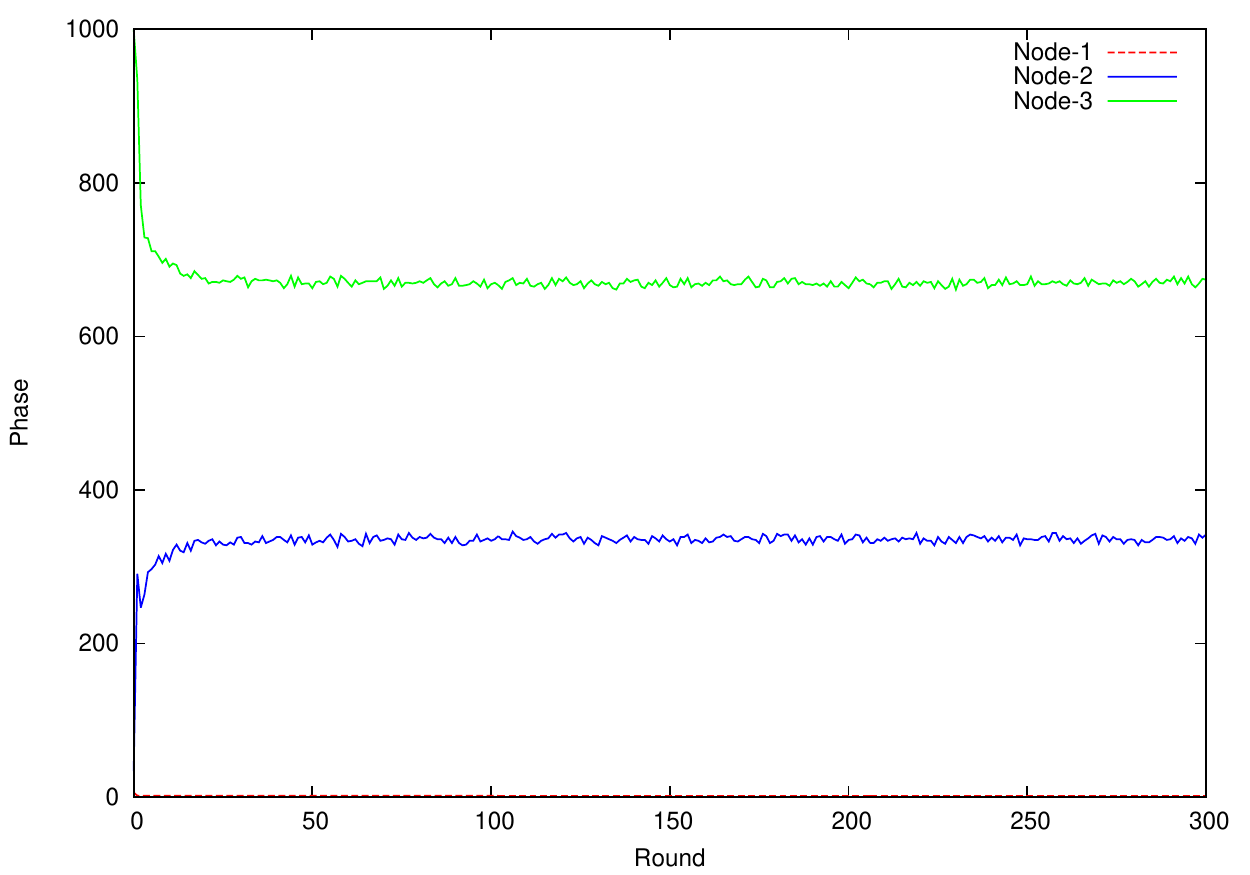}
		\label{fig:3nodes-chain-expected}
	}
	\caption{(a) shows noisy phases due to message collision at node 2. (b) shows the expected result of the perfect desynchrony state.}
\end{figure}

We simulate DWARF by setting the time period to 1000 milliseconds with nodes starting up randomly. 
The simulation result is shown in Figure \ref{fig:3nodes-chain-dwarf}. Node 2's and node 3's phases are plotted relatively to the node 1's phase which is relatively plotted to itself at 0. The noisy vertical line is the wrapping-around phase of node 3. It shows that node 1 and node 3 fire messages approximately at the same phase, causing message collision at node 2.
However, the expected result (\textit{i.e.} perfect desynchrony state) should be that three nodes are separated equivalently because all nodes will interfere each other if they fire messages at the same phase. The expected result is shown in Figure \ref{fig:3nodes-chain-expected} where each node is equivalently separated from each other approximately by 1000/3 milliseconds.

The problematic result is caused by the hidden terminal problem as demonstrated in Figure \ref{fig:3nodes-chain-hidden}; node 1 and node 3 are hidden to each other in this multi-hop topology. While node 3 is firing a message, node 1 senses a wireless channel and does not detect any signal from node 3 because the signal from node 3 is not strong enough within the signal range of node 1 and vice versa. Therefore, node 1 and node 3 notice only that there are two nodes, which are itself and node 2, in their perceived networks. Therefore, node 1 and node 3 simultaneously attempt to adjust their phases to the opposite side of node 2 in their time circles which are the same phase. As a result, messages of node 1 and 3 recursively collide at node 2. 

The hidden terminal problem does not only affect the performance of DWARF but also affect that of DESYNC.
This is due to the fact that, in DESYNC, a node adjusts its phase based on firing messages from its perceived phase neighbors. Therefore, if it fails to know the phase or presence of its two-hop neighbors, none of their phases is perceived. In \cite{MK09DESYNC}, EXT-DESYNC, an extension of DESYNC, is proposed to solve the hidden terminal problem based on a relative time relaying mechanism. Based on the similar idea, we extend DWARF to support multi-hop topologies.
However, only relative time relaying mechanism does not lead DWARF to an optimal solution in some cases. Therefore, this paper also proposes a \textit{force absorption} mechanism for extending DWARF to support multi-hop networks more efficiently.   

\subsubsection{Relative Time Relaying}
\label{sec:relative}
The first idea to solve the hidden terminal problem is straightforward. If a node does not know the firing times of its second-hop neighbors, its one-hop neighbors relay such information. Therefore, instead of firing only to notify others of its phase, each node includes their one-hop neighbors' firing time into a firing message. 

\begin{figure}
	\centering
	\subfloat[]{
		\label{fig:broadcast-problem-topo}
		\includegraphics[width=1.8in]{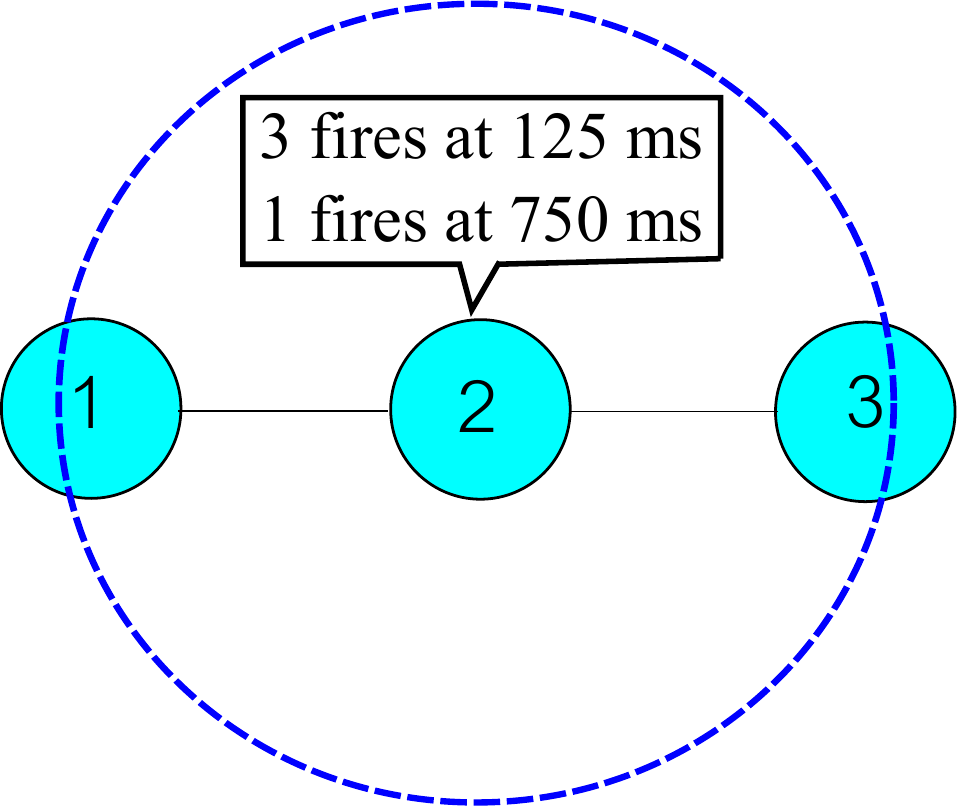}
	}
	\hspace{2cm}
	\subfloat[]{
		\label{fig:broadcast-problem-ring}
		\includegraphics[width=1.8in]{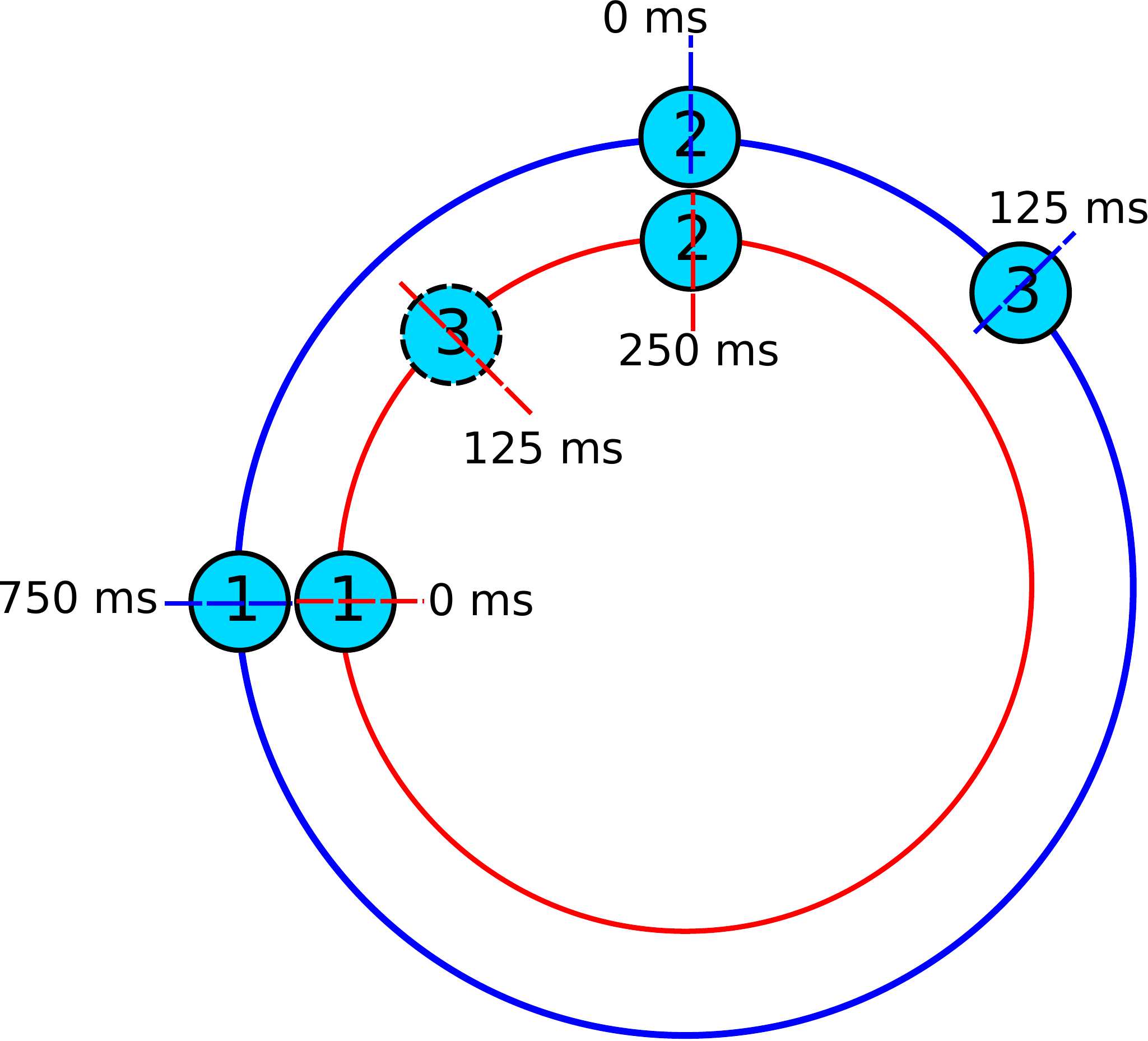}
	}
	\caption{A problem of relying on second-hop neighbors' firing timestamps from its one-hop neighbors. (a) shows node 2 firing a message. (b) displays node 1's misperception of phases.}
	\label{fig:broadcast-problem}
\end{figure}

However, due to our assumption that nodes' clocks are not synchronized, relying on second-hop neighbors' firing timestamps from its one-hop neighbors could lead to wrong phase adjustment. This problematic scenario is demonstrated in Figure \ref{fig:broadcast-problem}. Figure \ref{fig:broadcast-problem-topo} illustrates the firing message of node 2 that contains timestamps of its one-hop neighbors and Figure \ref{fig:broadcast-problem-ring} displays the problem. The inner circle represents the local time of node 1 and the outer circle represents the local time of node 2. The figure indicates that the local reference times (at 0 millisecond) of node 1 and node 2 are different. Therefore, if node 1 uses the node 3's firing time relayed by node 2, which is 125 milliseconds, node 1 will misunderstand the exact time phase of node 3. The misunderstood phase of node 3 is depicted as a dash circle. 

\begin{figure*}
	\centering
	\subfloat[]{
		\label{fig:broadcast-relative-topo}
		\includegraphics[width=1.8in]{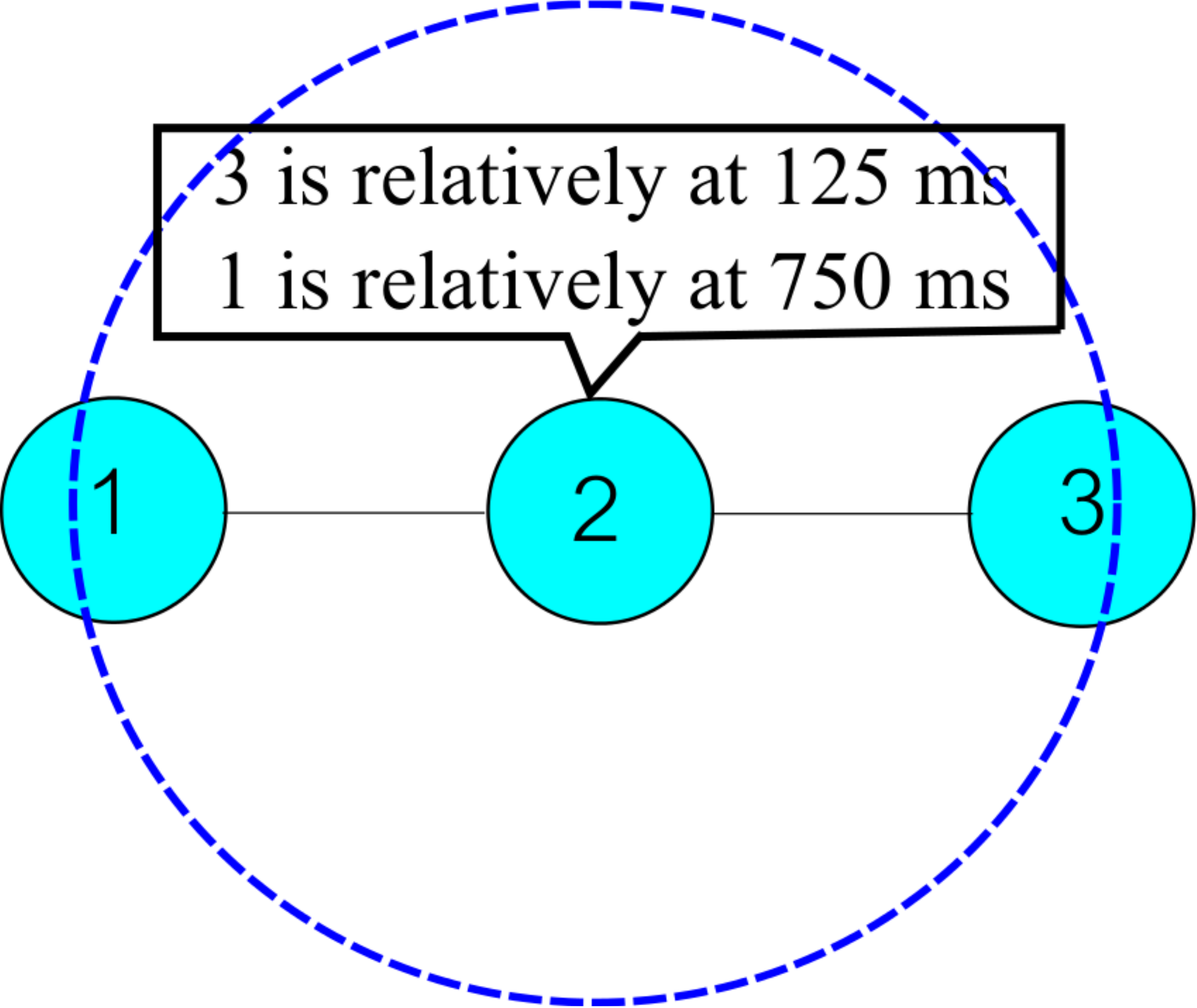}
	}
	\hfill
	\subfloat[]{
		\label{fig:broadcast-relative-ring}
		\includegraphics[width=1.8in]{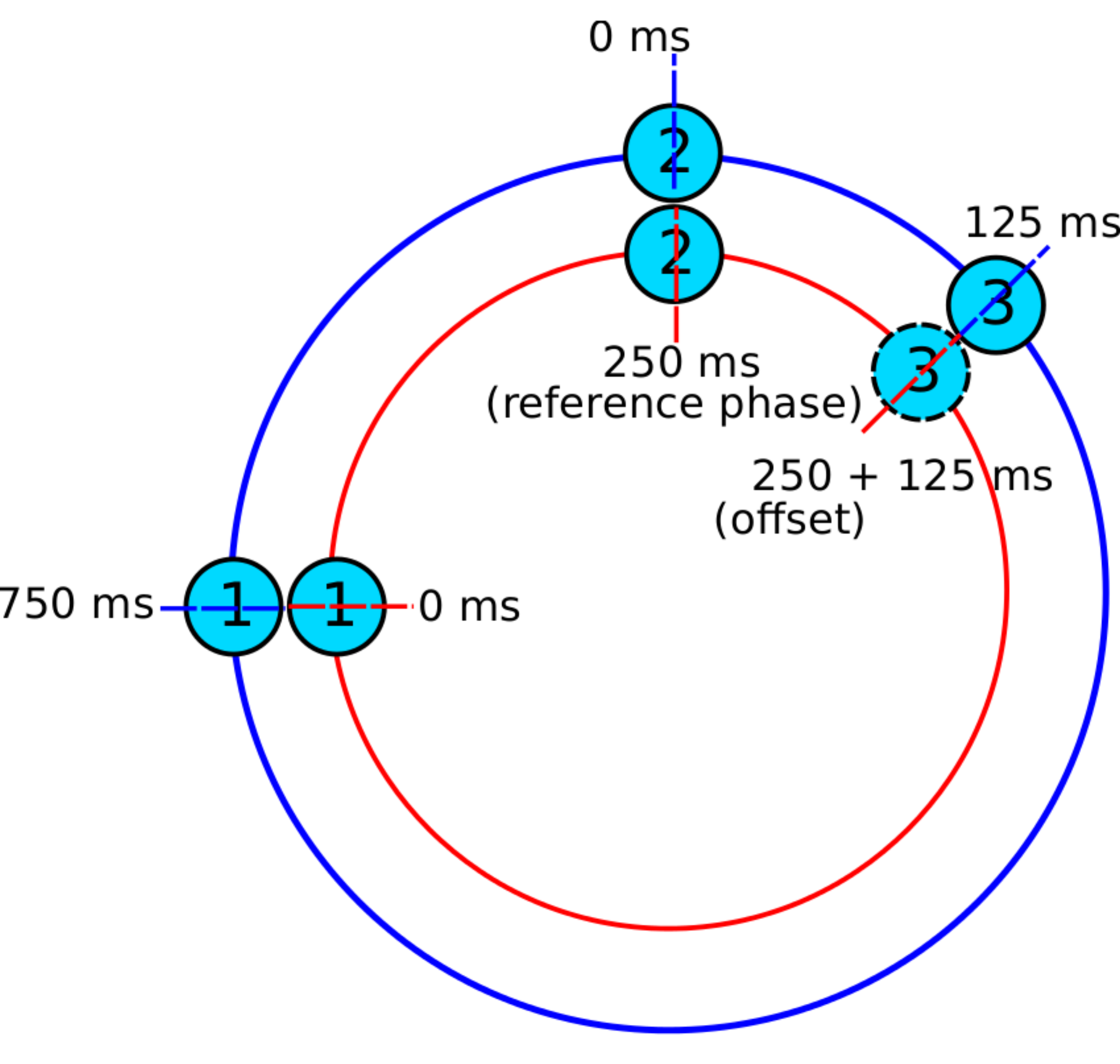}
	}
	\hfill
	\subfloat[]{
		\label{fig:broadcast-relative-perfect}
		\includegraphics[width=1.8in]{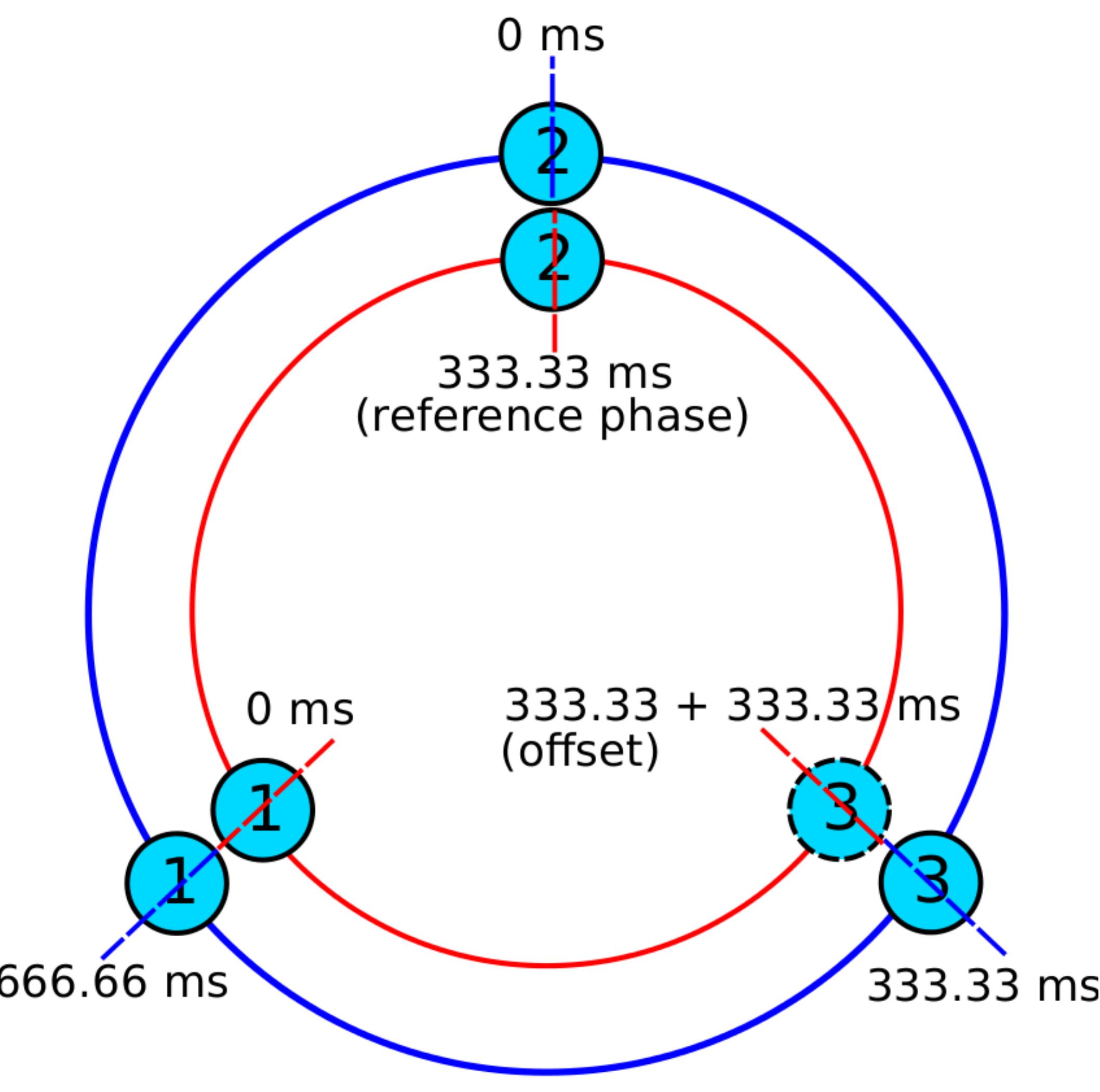}
	}
	\caption{M-DWARF solves the problem by using one-hop neighbors' relative phases. (a) shows node 2's firing message.  (b) shows how node 1 marks the node 2's phase as a reference phase and uses it as an offset for calculating the node 3's phase. (c) Eventually, nodes are in the perfect desynchrony state.}
	\label{fig:broadcast-relative}
\end{figure*}

This problem can be simply solved by using relative phases instead of actual local timestamps. Each node fires a message that includes relative phases of its one-hop neighbors. A receiving node marks the firing phase of the firing node as a reference phase. Then, the receiving node perceives its second-hop neighbors' phases as relative phase offsets to the reference phase. Figure \ref{fig:broadcast-relative} shows how M-DWARF desynchronizes a three-node multi-hop chain network.

\subsubsection{Force Absorption}
\label{sec:absorption}
\begin{figure*}
	\centerline{
		\subfloat[]{\includegraphics[scale=0.3]{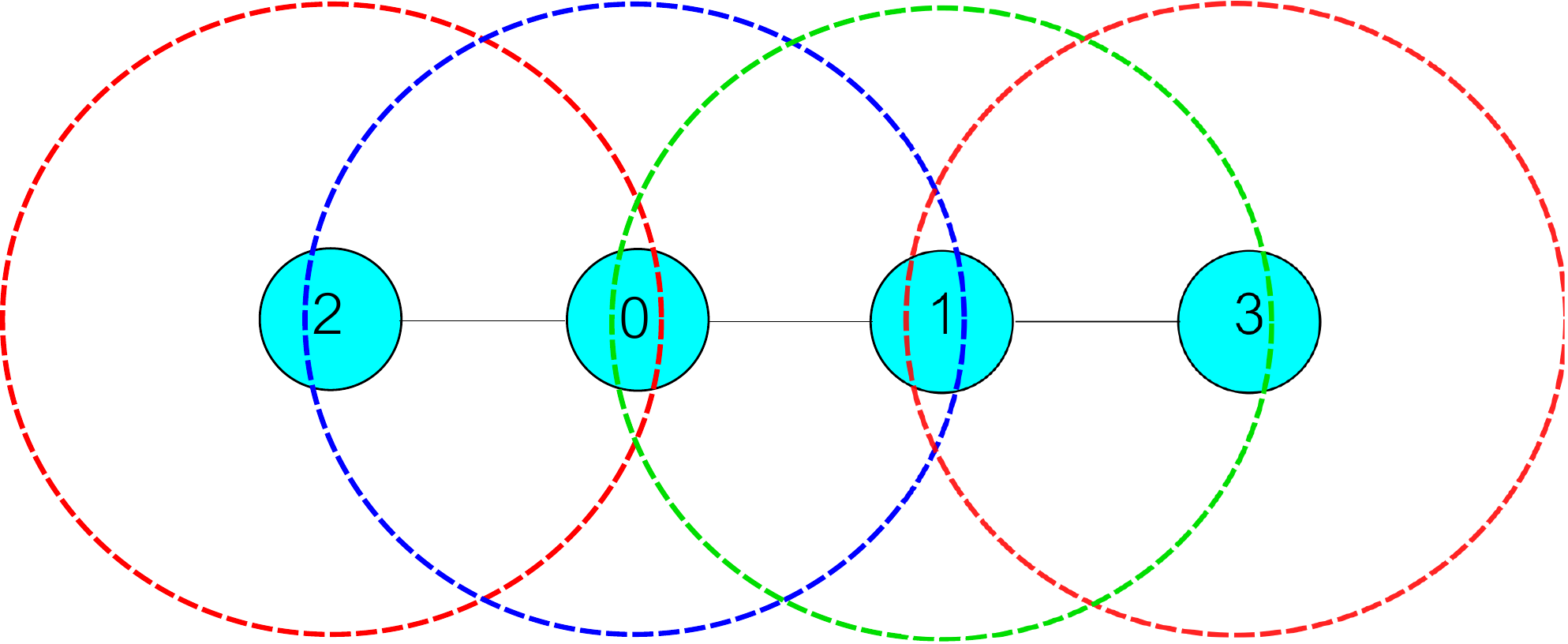}%
			\label{fig:4nodes-chain-topo}}
		\hfill
		\subfloat[]{\includegraphics[scale=0.20]{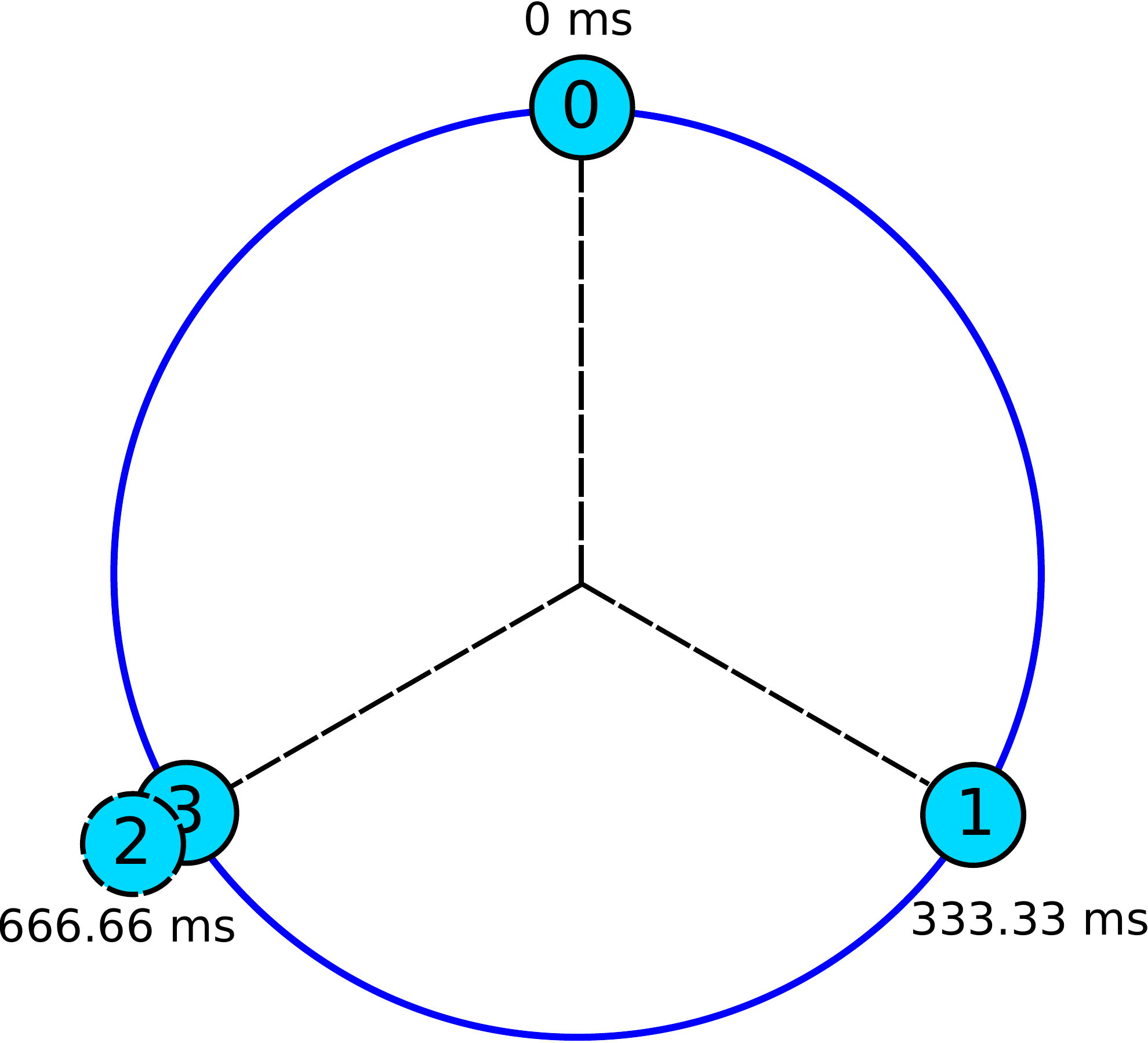}%
			\label{fig:4nodes-chain-expected}}
		\hfill
		\subfloat[]{\includegraphics[scale=0.20]{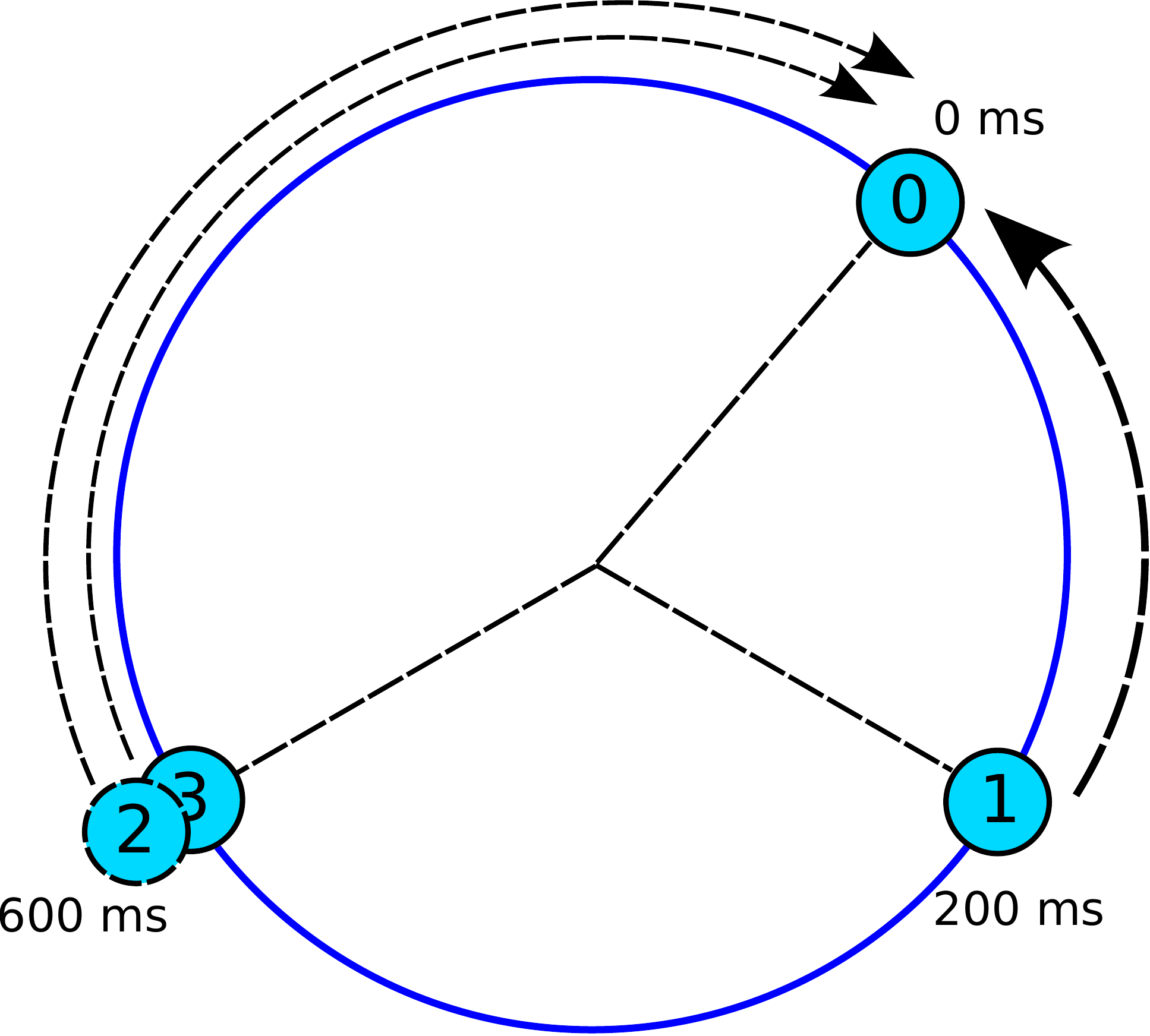}%
			\label{fig:4nodes-chain-dwarf}}
	}
	\caption{The problem of the DWARF algorithm. (a) displays 4-node chain topology. (b) shows node 0's local view. (c) displays an imperfect desynchrony state.}
	\label{fig:4nodes-chain}
\end{figure*}

As we mentioned earlier, DWARF with the relative time relaying mechanism does not solve some cases.
These cases are when there are at least two second-hop neighbors that can share the same phase without interference. For example, in a 4-node chain network illustrated in Figure \ref{fig:4nodes-chain-topo}, node 2 and node 3 are physically far beyond two hops. Therefore, they can fire messages at the same time phase without causing message collisions, as shown in Figure \ref{fig:4nodes-chain-expected}.
However, in M-DWARF, node 0 perceives that node 2 and node 3 are at the same phase. Therefore, there are two forces from node 2 and node 3 to repel node 0 clockwise but there is only force from node 1 to repel node 0 counter-clockwise. Consequently, node 0 cannot stay at the middle between node 1 and the group of node 2 and 3 (see Figure \ref{fig:4nodes-chain-dwarf}). 

Therefore, we propose a novel force absorption mechanism for multi-hop desynchronization based on the artificial force field. The objective of this mechanism is to absorb the overwhelming force from at least two nodes that can fire at the same phase without interference.

\begin{figure*}
	\centerline{
		\subfloat[]{\includegraphics[scale=0.2]{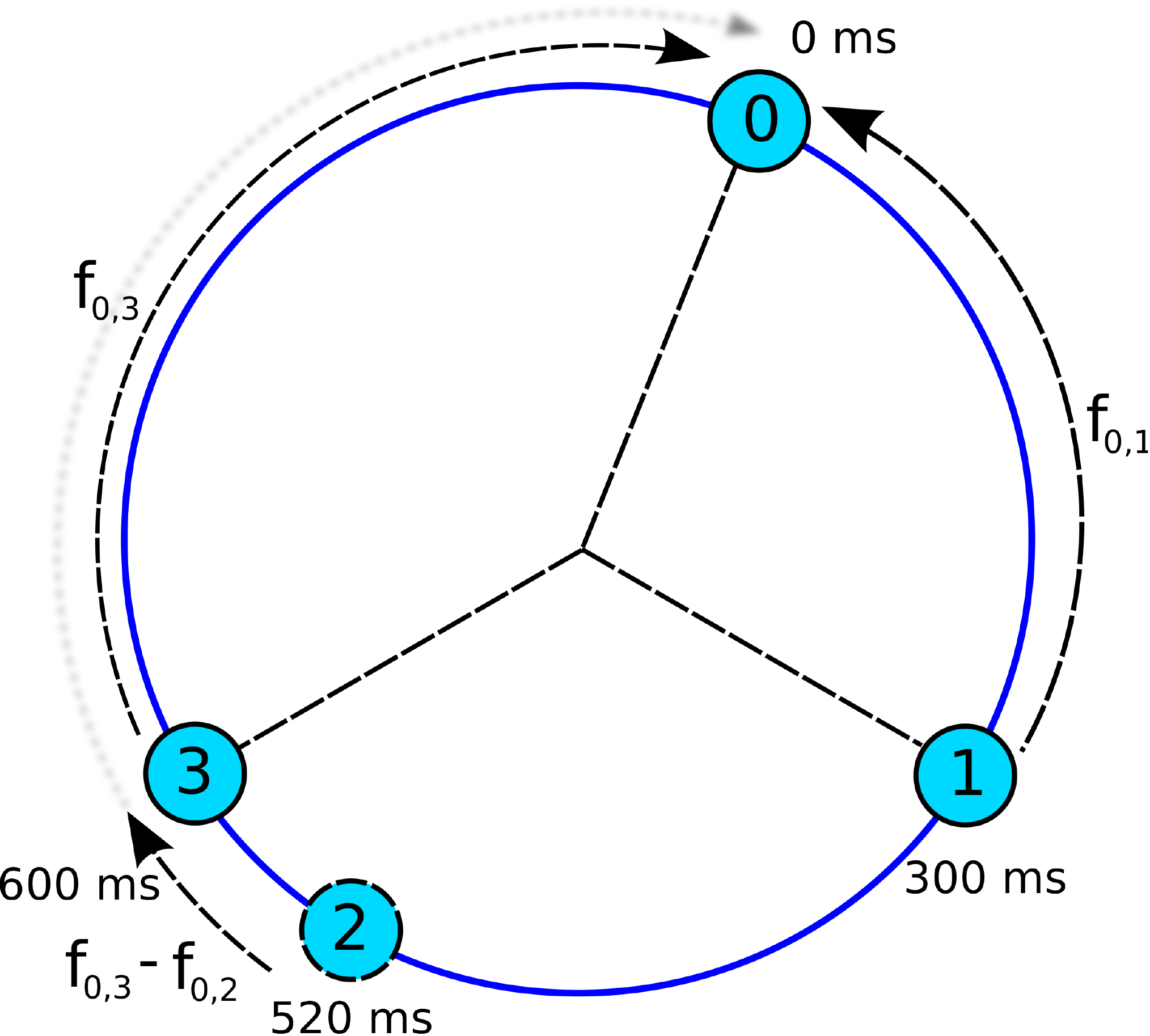}%
			\label{fig:4nodes-chain-dwarf-absorb-split}}
		\hfil
		\subfloat[]{\includegraphics[scale=0.2]{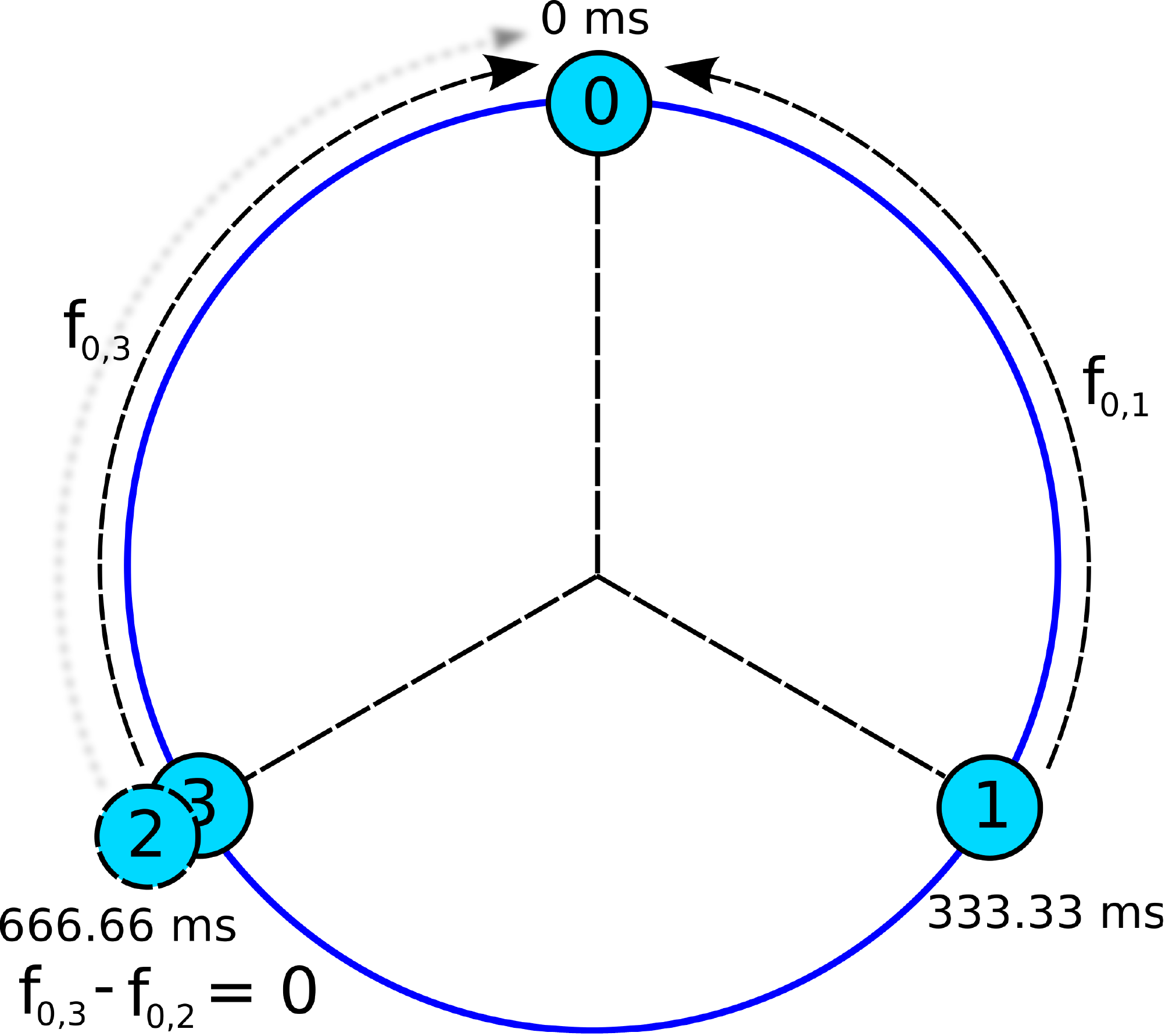}%
			\label{fig:4nodes-chain-dwarf-absorb-perfect}}
	}
	\caption{M-DWARF solves the problem with force absorption; the blur line represented an absorbed force. (a) shows node 2's force is being absorbed by node 3. (b) displays the perfect desynchrony state.}
	\label{fig:4nodes-chain-dwarf-absorb}
\end{figure*}

The mechanism works as follows. A node receives a full repulsive force from the next/previous phase neighbor as in DWARF. However, a force from the second-next / second-previous phase neighbor is partially absorbed by the next / previous phase neighbor. The magnitude of the absorbed force depends on the phase interval between the next / previous and the second-next / second-previous phase neighbors.  The closer the second-next / second-previous phase neighbor moves to the next / previous phase neighbor, the lower the magnitude of the absorbed force becomes. Eventually, when the second-next / second-previous phase neighbor moves to the same phase as the next / previous phase neighbor, the additional force from the second-next / second-previous phase neighbor is fully absorbed. Consequently, the magnitude of two forces repelling the considered node is approximately equal to only the magnitude of one force. This principle is applied recursively; that is, the force from the third-next / third-previous phase neighbors is absorbed by the second-next / second-previous phase neighbor, and the force from the fourth-next / fourth-previous phase neighbor is absorbed by the third-next/third-previous phase neighbor, and so forth.   
Figure \ref{fig:4nodes-chain-dwarf-absorb} illustrates this mechanism. In Figure \ref{fig:4nodes-chain-dwarf-absorb-split}, the force from node 2 to node 0 is absorbed by node 3 (the absorbed force is displayed in a blur line). Thus, from node 2, there is only small magnitude of force left to node 0. Eventually, in Figure \ref{fig:4nodes-chain-dwarf-absorb-perfect}, node 2 moves to the same phase as node 3 because they do not interfere each other and the force from node 2 is fully absorbed. Consequently, the network can be in the perfect desynchrony state.

Let $f_{i,j}$ be a full repulsive force from node $j$ to node $i$, $f_{i,j}^{'}$ be an absorbed force from node $j$ to node $i$, $T$ is the time period, and $\Delta \phi_{i,j}$ is the phase difference between node $i$ and $j$. 
The force function for multi-hop networks is the following:
\begin{alignat}{2}
f_{i,j} &= \frac{1}{\Delta \phi_{i,j} / T}, \text{where }\Delta \phi_{i,j} \in (-\frac{T}{2}, \frac{T}{2}) \nonumber \\
f_{i,i + 1}^{'} &= f_{i,i + 1} \nonumber \\
f_{i, i - 1}^{'} &= f_{i,i - 1} \nonumber \\
f_{i,j}^{'} &= f_{i,x} - f_{i,j}, 
\label{eq:force-absorb}
\end{alignat}
where $j \notin \left\{i -1, i + 1\right\}$ and $x = (j - \frac{\Delta \phi_{i,j}}{|\Delta \phi_{i,j}|}) \mod n$.
For $f_{i,x}$, if node $j$ repels node $i$ forward, $x$ is $j + 1$. In contrast, if node $j$ repels node $i$ backward, $x$ is $j - 1$. At $T/2$ or $-T/2$, a node does not repel an opposite node because they are balanced.

For example, in Figure \ref{fig:4nodes-chain-dwarf-absorb}, node 0 calculates the force from node 2 as the following:
\begin{alignat}{2}
f_{0,2}^{'} &= f_{0,3} - f_{0,2} \nonumber \\
&= \frac{1}{\Delta \phi_{0,3} / T} - \frac{1}{\Delta \phi_{0,2} / T}. \nonumber
\end{alignat} 

Noticeably, if node 2 moves close to node 3, the value of $\Delta \phi_{0,2}$ is close to the value of $\Delta \phi_{0,3}$. Then, the magnitude of force $f_{0,2}$ is reduced. 
Finally, when $\Delta \phi_{0,2}$ is equal to $\Delta \phi_{0,3}$ as in Figure \ref{fig:4nodes-chain-dwarf-absorb-perfect}, the magnitude of force $f_{0,2}$ becomes 0; in other words, the force is fully absorbed.

\newpage
\appendix
\section*{APPENDIX}
\section{Psuedocodes of M-DWARF}
\label{sec:psuedocode_m_dwarf}
\begin{algorithmic}[1]
	\STATE \textbf{Initialization}  
	\STATE $T = TimePeriod$ \COMMENT{Configurable Time Period}
	\STATE $n = 1$ \COMMENT{Number of neighbors within two hops including itself}
	\STATE $\mathcal{F} = 0$ \COMMENT{Force Summation}
	\STATE $phasesBuffer = Array$ \COMMENT{Buffer for phases of neighbors within two hops}
	\STATE $lastFiringTime = localTime$
	\STATE $currentPhase = localTime$ modulo $T$
	\STATE Set a firing timer to be $T$ unit time
	\newline
	
	\STATE \textbf{Upon timer firing}
	\STATE Broadcast a firing message containing relative phases of one-hop neighbors
	\STATE $lastFiringTime = localTime$
	\STATE $currentPhase= localTime$ modulo $T$
	\STATE $K = 38.597 \times n^{-1.874} \times \frac{T}{1000}$
	\STATE $\mathcal{F}$ = call Calculate total force
	\STATE $newPhase = currentPhase + (K \times \mathcal{F})$
	\IF{$newPhase < 0$}
	\STATE $newPhase = T + newPhase$
	\ENDIF
	\STATE Set a firing timer to be fired at ($newPhase$ modulo $T$)
	\STATE $\mathcal{F} = 0$
	\STATE $n = 1$
	\newline
	
	\STATE \textbf{Calculate total force}
	\STATE $sortedBuffer$ = Sort $phasesBuffer$
	\STATE $forwardForce = \frac{1}{(T - sortedBuffer[length - 1])/T}$
	\STATE $backwardForce = \frac{1}{sortedBuffer[0]/T}$
	\FOR{$i$ in range of 1 to length of sorted $phasesBuffer$ - 2}
	\IF{$phasesBuffer[i] > 0.5T$}
	\STATE $forwardForce = forwardForce + \newline (\frac{1}{(T - phasesBuffer[i+1]) / T} - \frac{1}{(T - phasesBuffer[i]) / T})$
	\ELSE
	\STATE $backwardForce = backwardForce + \newline (\frac{1}{phaseBuffer[i-1] / T} - \frac{1}{phaseBuffer[i] / T})$
	\ENDIF
	\ENDFOR
	\STATE return $forwardForce - backwardForce$
	\newline
	
	\STATE \textbf{Upon receiving a firing message from $nodeId$}
	\STATE $phaseDiff = localTime - lastFiringTime$
	\STATE $phasesBuffer[nodeId] = phaseDiff$
	\STATE $reference = phaseDiff$
	\FOR{each $twoHopNodeId$ of two-hop neighbors in $message$}
	\IF{$twoHopNodeId$ is not in $phasesBuffer$} 
	\STATE $phasesBuffer[twoHopNodeId] = \newline message[twoHopNodeId][relativePhase] + reference$
	\STATE $n = n + 1$
	\ENDIF
	\ENDFOR
\end{algorithmic}

\newpage
\bibliographystyle{unsrt}      

\bibliography{paper}   % name your BibTeX data base

\end{document}